\documentclass[aps, prd, preprint, superscriptaddress,  preprintnumbers, showkeys,showpacs, nofootinbib,floatfix ]{revtex4}
\usepackage[dvips]{graphicx}
\usepackage{latexsym}
\usepackage{natbib}
\usepackage{amssymb}
\usepackage{amsmath}
\usepackage{epstopdf}
\usepackage{booktabs}
\usepackage{dcolumn}

\newcolumntype{.}{D{.}{.}{-1}}

\DeclareGraphicsExtensions{.eps}

\begin{document}
\title{
  Finite temperature study of the axial $U(1)$ symmetry on the lattice
  with overlap fermion formulation
} 
\date{\today}

\author{Guido Cossu}
\affiliation{Theory Center, IPNS, High Energy Accelerator Research Organization (KEK), Tsukuba 305-0810, Japan}
\author{Sinya Aoki}
\affiliation{Yukawa Institute for Theoretical Physics, Kyoto University, Kitashirakawa Oiwakecho, Sakyo-ku, Kyoto 606-8502, Japan}
\affiliation{Center for Computational Sciences, University of Tsukuba, Tsukuba, Ibaraki 305-8577, Japan}
\author{Hidenori Fukaya}
\affiliation{Department of Physics, Osaka University, Toyonaka 560-0043, Japan}
\author{Shoji Hashimoto}
\affiliation{Theory Center, IPNS, High Energy Accelerator Research Organization (KEK), Tsukuba 305-0810, Japan}
\affiliation{School of High Energy Accelerator Science, The Graduate University for Advanced Studies (Sokendai), Tsukuba 305-0801, Japan}
\author{Takashi Kaneko}
\affiliation{Theory Center, IPNS, High Energy Accelerator Research Organization (KEK), Tsukuba 305-0810, Japan}
\affiliation{School of High Energy Accelerator Science, The Graduate University for Advanced Studies (Sokendai), Tsukuba 305-0801, Japan}
\author{Hideo Matsufuru}
\affiliation{Theory Center, IPNS, High Energy Accelerator Research Organization (KEK), Tsukuba 305-0810, Japan}
\author{Jun-Ichi Noaki}
\affiliation{Theory Center, IPNS, High Energy Accelerator Research Organization (KEK), Tsukuba 305-0810, Japan}

\begin{abstract}
  We examine the axial $U(1)$ symmetry near and above the
  finite temperature phase transition in two-flavor QCD using lattice
  QCD simulations.
  Although the axial $U(1)$ symmetry is always violated by quantization, 
  {\it i.e.} the chiral anomaly, the correlation functions may manifest
  effective restoration of the symmetry in the high temperature phase.
  We explicitly study this possibility by calculating the meson
  correlators as well as the Dirac operator spectral density near
  the critical point.
  Our numerical simulations are performed on a $16^3\times 8$ lattice
  with two flavors of dynamical quarks represented by the overlap
  fermion formalism.
  Chiral symmetry and its violation due to the axial anomaly is
  manifestly realized with this formulation, which is a prerequisite
  for the study of the effective restoration of the axial $U(1)$
  symmetry. 
  In order to avoid discontinuity in the gauge configuration space,
  which occurs for the exactly chiral lattice fermions,
  the simulation is confined in a fixed topological sector.
  It induces finite volume effect, which is well described by a
  formula based on the Fourier transform from the $\theta$-vacua.
  We confirm this formula at finite temperature by calculating the
  topological susceptibility in the quenched theory. Our two flavor 
  simulations show degeneracy of the meson correlators and a gap in the 
  Dirac operator spectral density, which implies that the axial $U(1)$ symmetry
  is effectively restored in the chirally symmetric phase.
\end{abstract}
\pacs{}
\keywords{Lattice QCD, Topology, overlap-Dirac operator, U(1), symmetry}
\preprint{KEK-CP-285}
\preprint{OU-HET-771-2012}
\preprint{UTHEP-656}
\preprint{YITP-13-24}

\maketitle

\section{Introduction}
Chiral symmetry plays a key role in understanding the nature of
the vacuum of Quantum Chromodynamics (QCD). 
With $N_f$ flavors of massless quarks, the QCD lagrangian has a symmetry 
$U(N_f)_L\otimes U(N_f)_R$ under field rotations involving flavor
mixings for left-handed and right-handed quarks independently.
Among them the flavor-singlet part of the chiral (or axial)
transformation, which forms $U(1)_A$, is violated by quantization.
This is known as the chiral or axial anomaly. 
At zero temperature, the remaining symmetry 
$SU(N_f)_L\otimes SU(N_f)_R\otimes U(1)_V$ 
is spontaneously broken down to $SU(N_f)_V\otimes U(1)_V$,
and $N_f^2-1$ massless Nambu-Goldstone bosons appear.
At finite temperature, in the massless limit, a restoration of symmetry is expected back to 
$SU(N_f)_L\otimes SU(N_f)_R\otimes U(1)_V$ 
above a critical temperature $T_c$.

This symmetry breaking and its restoration are characterized by a
vacuum expectation value of a flavor-singlet scalar operator
$\bar{q}q$, which is called the chiral condensate
$-\langle\bar{q}q\rangle\equiv\Sigma$, the order parameter of this phase transition. 
According to the Banks-Casher relation \cite{Banks:1979yr}
\begin{equation}
  \label{eq:Banks-Casher}
  \Sigma = \pi\rho(0),
\end{equation}
non-zero chiral condensate implies an accumulation of low-lying
eigenvalues of the Dirac operator $D$.
Here, $\rho(\lambda)$ is a spectral density of $D$ defined as
$\rho(\lambda)=(1/V)\sum_{\lambda'}\langle\delta(\lambda-\lambda')\rangle$
with $V$ a four-volume of space-time and 
$\langle\cdots\rangle$ denotes an expectation value.
The Banks-Casher relation (\ref{eq:Banks-Casher}) is satisfied in the
thermodynamical limit, {\it i.e.}
infinite volume limit, followed by the limit of vanishing quark mass.
This implies that the density of low-lying eigenvalues of $D$ must disappear
above the transition temperature.

The chiral anomaly is reflected in the particle spectrum (in the low temperature
phase) as a mass-splitting of the corresponding pseudo-scalar
meson, the $\eta$ meson in the case of $N_f=2$ for example, from the flavor
non-singlet pseudo-scalar particles, {\it i.e.} $\pi$ mesons.
In terms of (valence) quark flow diagrams, the difference between
flavor-singlet and non-singlet mesons comes from a disconnected
diagram in the meson two-point correlation functions. It is found that 
the main contribution to the disconnected diagram is
from low-lying quark eigenmodes if one decomposes the quark propagator
into the contributions of individual eigenmodes of the Dirac operator
\cite{Aoki:2007pw}.
An interesting question then emerges:
above the finite temperature phase transition, where we expect the
suppression (or even absence) of the low-lying eigenmodes, 
does $\eta$ become degenerate with flavor non-singlet pseudo-scalars?
If so, the axial $U(1)$ symmetry is effectively restored at least in the
particle spectrum at high temperature.
This may have significant impact in the nature of the finite
temperature phase transition, since the symmetry controls the order
and critical exponents of the phase transition \cite{Pisarski:1983ms}.

Such argument was originally made by Cohen \cite{Cohen:1996ng},
followed by several theoretical studies 
\cite{Evans:1996wf,Lee:1996zy,Birse:1996dx,Schafer:1996hv,Aoki:2012yj}.
Numerical study using lattice QCD, on the other hand, has been
missing, because most numerical simulations at finite temperature
performed so far 
used the staggered fermion formulation, which realizes chiral
symmetry only partially.
Even more important problem of the staggered fermion is in the
so-called {\it rooting} procedure, {\it i.e.}
a square-root of the fermion determinant is taken to represent two
flavors of light quarks in the nature out of four artificial flavors
built in the staggered fermion.
The flavor-singlet axial transformation is ill-defined in
such a theory. There is a recent study of the Dirac spectral density
using the staggered fermion \cite{Ohno:2012br}.
In this paper, we present the first numerical study of this problem
employing the overlap fermion formulation 
\cite{Neuberger:1997fp,Neuberger:1998wv}, for which the full flavor
and chiral symmetries are realized on the lattice including the
flavor-singlet axial symmetry existing at the Lagrangian level and
violated by quantization.

Although the numerical cost for the simulation of the overlap fermion is
substantially higher than that of the staggered fermion, its simulation
has become feasible by recent developments of machines and algorithms,
and in fact has been extensively performed by the JLQCD and TWQCD
collaborations in the past years \cite{Aoki:2012pma}. 
The overlap fermion is constructed with the overlap-Dirac operator
$D_{\rm ov}$,
\begin{equation}
  D_{\rm ov} = m_0[1 + \gamma_5 {\rm sgn}(H_W(-m_0))],
  \label{eq:Dov}
\end{equation}
where $H_W(-m_0)$ is the hermitian Wilson-Dirac operator
$H_W(-m_0)=\gamma_5D_W(-m_0)$ with a large negative mass $-m_0$,
which is of order of the lattice cutoff.
This operator satisfies the Ginsparg-Wilson relation
\cite{Ginsparg:1981bj},
$D_{\rm ov}\gamma_5+\gamma_5D_{\rm ov}=D_{\rm ov}\gamma_5D_{\rm ov}/m_0$,
through which the exact chiral symmetry on the lattice can be defined
\cite{Luscher:1998pqa}.
The operator (\ref{eq:Dov}) has a singularity due to the sign
function, when an eigenvalue of $H_W(-m_0)$ crosses zero as the
background gauge field varies. 
This corresponds to a boundary between two adjacent topological
sectors of gauge field.
Numerical simulation with currently available algorithms becomes
prohibitively costly when one crosses this singularity;
the JLQCD and TWQCD collaborations took a strategy to fix the
topological sector during the Monte Carlo simulation.
Physical quantities at the $\theta=0$ vacuum is reconstructed later by
correcting the finite volume effect of $O(1/V)$ due to fixing the
topology \cite{Brower:2003yx,Aoki:2007ka}.

Since the artifact due to fixing topology is essentially a finite
volume effect, the same strategy should work at finite temperature as
far as the spatial volume is sufficiently large.
Detailed discussions are given in Section~\ref{sec:fixed_Q}.
We numerically check this property by calculating the topological
susceptibility on quenched lattices, which is given in
Section~\ref{sec:pure_gauge}. 
As discussed in \cite{Aoki:2007ka}, even when the global topology is fixed,
local topological fluctuations are still active and the topological
susceptibility can be extracted from long-range correlation of
topological charge densities.
The results at finite temperature obtained in this work
are compared with conventional calculations from the
fluctuation of global topology found in the literature.

We then discuss the results of our exploratory simulations of 
two-flavor QCD at finite temperature 
using the overlap fermion formulation in Section~\ref{sec:TwoFlav}. 
The global topological charge is fixed to zero.
We analyze the low-lying eigenvalue spectral density of the overlap-Dirac
operator,
as well as the meson correlators to investigate the effective
restoration of the axial $U(1)$ symmetry.
Both connected and disconnected parts of the correlator are reconstructed 
from the low-lying eigenmodes, which are calculated and stored in advance. 
The disconnected part produces the difference between the meson
channels related by the $U_A(1)$ symmetry transformations. 
We show that, in the high temperature phase, the disconnected
contribution is indeed vanishing towards the chiral limit of sea
quarks. 

A similar study of the restoration of the axial $U(1)$ symmetry has
recently been made by the HotQCD collaboration \cite{:2012jaa} using
the domain-wall fermion formalism.
Like the overlap fermion, the domain-wall fermion realizes exact
chiral symmetry on the lattice, but only in the limit of infinitely
large fifth dimension.
In practical simulations, the chiral symmetry is slightly violated
and an additive mass renormalization $m_{res}$ appears.
The size of $m_{res}$ is typically an order of MeV, and poses a
significant problem when one tries to identify the near-zero eigenvalues,
which are of the same order or even lower.
The overlap fermion employed in this work enables a clear
identification of the near-zero modes and their effects on the
physical quantities.

Reports of our work at earlier stages are found in
\cite{Cossu:2012gm,Cossu:2012jj}.

\section{Physics at fixed topology}
\label{sec:fixed_Q}

In QCD, the fluctuation of global topology $Q$ is necessary to
guarantee the cluster decomposition property, which is one of the
fundamental conditions necessary in constructing meaningful quantum
field theory. 
It is easy to see that the different topological sectors have to be
added with a weight of the form $e^{i\theta Q}$, through which the QCD
$\theta$-parameter is defined \cite{Weinberg:1996kr}. 
The formulae to relate the physical quantities obtained in a fixed
topological sector $Q$ to those in the $\theta=0$ vacuum are developed in 
\cite{Brower:2003yx,Aoki:2007ka}.
They are valid also at finite temperature as outlined below.

Let $Z(\theta)$ be a partition function of QCD with the $\theta$-term
put in a volume $V=L^3\times\beta$.
Here, $L$ and $\beta$ are spatial and temporal extent of the box
respectively ($\beta = N_ta$).
The partition function $Z_Q$ at a fixed global topological charge $Q$
can be written as a Fourier transform of the $\theta$-vacua: 
\begin{equation}
  \label{eq:Z_Q}
  Z_Q =
  \frac{1}{2\pi}\int_{-\pi}^{\pi}\!d\theta\, Z(\theta) \exp(i\theta Q)
  =
  \frac{1}{2\pi}\int_{-\pi}^{\pi}\!d\theta\,\exp( -\beta F(\theta)L^3),
\end{equation}
where 
$F(\theta)\equiv E_0(\theta)-i\theta Q/V$.

Near zero temperature, $E_0(\theta)$ represents the vacuum energy density,
$Z(\theta) = \exp(-\beta E_0(\theta) L^3)$.
The integral in (\ref{eq:Z_Q}) may be evaluated using the saddle
point expansion around
\begin{equation}
  \theta_c = i \frac{Q}{\chi_t V}(1+O(\delta^2)),
\end{equation}
where $\chi_t$ is defined through an expansion of 
$E_0(\theta)$ in terms of $\theta$,
\begin{equation}
  E_0(\theta) = 
  \sum_{k=1}^\infty  \frac{c_{2k}}{(2k)!} \theta^{2k} 
  = \frac{\chi_t}{2}\theta^2 + O(\theta^4),
\end{equation}
and $\delta\equiv Q/(\chi_t V)$.
The result up to small $\delta^2$ terms is given by
\begin{equation}
  \label{eq:saddle_point Z_Q}
  Z_Q = 
  \frac{1}{\sqrt{2\pi\chi_t V}}
  \exp\left(-\frac{Q^2}{2\chi_t V}\right) \times
  \left[
    1 -\frac{c_4}{8 V \chi_t^2} 
    + O\left(\frac{1}{V^{2}}, \delta^2 \right)
  \right].
\end{equation}
The global topological charge distributes as Gaussian with a width
specified by the topological susceptibility $\chi_t$.
At finite volume, the correction appears at order $1/V$.
The basic assumption in obtaining (\ref{eq:saddle_point Z_Q}) through
the saddle point expansion is that $E_0(\theta)$ takes its minimum
value at $\theta=0$ and is analytic around there.

At non-zero temperature, $\beta$ plays the role of inverse temperature.
As the first excited state energy $E_1(\theta)$ becomes significant
compared to the temperature,
$\beta(E_1(\theta)-E_0(\theta))L^3\sim 1$,
the partition function $Z(\theta)$ receives contributions from the
excited states as
$Z(\theta)=\sum_n\exp(-\beta E_n(\theta)L^3)$,
and the relation (\ref{eq:Z_Q}) has to be modified to include them.
This can be done by redefining the ``energy'' as 
$\beta\tilde{E}(\theta)L^3\equiv -\ln Z(\theta)$
and accordingly the ``free energy''
$\beta\tilde{F}(\theta)L^3\equiv\beta\tilde{E}(\theta)L^3-i\theta Q/V$.
As far as the analyticity property of $\tilde{E}(\theta)$ around
$\theta=0$ is unchanged, the same saddle-point approximation can be
applied, and
(\ref{eq:saddle_point Z_Q}) is valid with $\chi_t$ and $c_4$ 
evaluated at finite temperature. 
The condition that $1/V^2$ and $\delta^2$ have to be small
in (\ref{eq:saddle_point Z_Q}) is still applied.
Since $\beta$ is fixed, the condition implies large spatial volume
$L^3$ to make $V=\beta L^3$ sufficiently large.

The same conclusion is obtained by considering the ``energy'' defined
through a transfer matrix in one of three spatial directions.
In this way, the spatial extent $L$ plays the role of inverse
temperature and the whole argument given at vanishing temperature
remains unchanged.
Again, the volume $V$ appearing in the formulae is the space-time
volume $\beta L^3$.  

Along the same line of argument, one can obtain the relation between
correlation functions at fixed $Q$ and those at fixed $\theta$.
For instance, for a correlator $G(\theta)$ that is CP-even at
$\theta=0$, one obtains \cite{Aoki:2007ka}
\begin{eqnarray}
  G_Q
  &=& 
  G(0) + G^{(2)}(0)
  \frac{1}{2\chi_tV}
  \left[1-\frac{Q^2}{\chi_tV}-\frac{c_4}{2\chi_t^2V}\right]
  \nonumber \\
  && + G^{(4)}(0)\frac{1}{8\chi_t^2 V^2} + O(V^{-3}).
  \label{eq:even}
\end{eqnarray}
Namely, the correlator at a fixed topology $G_Q$ is written in terms
of $G(0)$ and its second derivative 
$G^{(2)}(0)\equiv dG(\theta)/d\theta|_{\theta=0}$, and so on.
The first correction is again of the order of $1/V$.

One important example of (\ref{eq:even}) is that for a two-point
correlation function of the topological charge density $\omega(x)$.
It can also be written in terms of flavor-singlet pseudo-scalar
density operators $mP(x)$ using the flavor-singlet
axial-Ward-Takahashi identity.
Fixing the global topology, a constant correlation remains at long
distances: 
\begin{equation}
  \lim_{\vert x\vert\to \mathrm{large}}
  \langle m P(x) m P(0) \rangle_Q =
  \frac{1}{V} \left(
    \frac{Q^2}{V}-\chi_t - \frac{c_4}{2 \chi_t V}
  \right)
  + O(e^{-m_{\eta}\vert x\vert}).
  \label{eq:singlet}
\end{equation}
At finite temperature, the long distance $|x|$ must be taken in the
spatial direction.

The constant correlation appearing in (\ref{eq:singlet}) is understood
as the effect of fixing topology. 
Let us consider a sector of $Q=0$, as an example, {\it i.e.} the
global topology is constrained to zero.
If there is a positive topological charge fluctuation locally near the
origin, there would be more chance to find negative fluctuation apart
from there in order that $\langle mP(x)\rangle$ is summed up to zero
when integrated over space-time. 
The correlation $\langle mP(x)mP(0)\rangle$ is thus negative at long
distances and is proportional to the ability of having local
topological fluctuations, which is characterized by $\chi_t$.
Since the effect must vanish on large enough volume, one expects a
contribution of the form $-\chi_t/V$,

The relation (\ref{eq:singlet}) suggests a possibility to extract
$\chi_t$ from the measurement done in the fixed topology.
Indeed, it was successfully performed at zero temperature in
\cite{Aoki:2007pw}.

The last term in (\ref{eq:singlet}) represents a physical correlation
due to the flavor-singlet pseudo-scalar particle, here denoted as $\eta$.
In the lattice calculation of the flavor-singlet correlation
functions, there appear connected and disconnected quark-flow diagrams.
Both diagrams contain a slow decay mode due to the flavor non-singlet
pion, which cancels between the two diagrams, and only the rapidly
decaying channel of heavier $\eta$ particle remains.
In the quenched theory, there is no pion component in the disconnected
sector and a modified contribution from the so-called hairpin diagram
has to be considered, as discussed later in
Section~\ref{sec:Diconnected_correlation}.

When $\chi_t$ is very small, which is expected for $N_f =2$ QCD above $T_c$,
our method for extracting $\chi_t$ is no longer valid.
However,  the disconnected propagator is still useful since
its absence (or presence) itself is a signal for
the $U_A(1)$ restoration (or breaking).
If there is a sizable gap in the Dirac spectral density,
it is likely that the disconnected diagram is highly suppressed and $\chi_t=0$. 
Conversely, if the eigenvalue density is non-zero
but very tiny, $\chi_t$ is likely to have a non-zero value.
In this way, even when $\chi_t$ is small, we can indirectly
investigate whether  $\chi_t =0$ or not, and $U_A(1)$ is restored or not,
through the fixed-topology simulations.

\section{Study of fixed topology in quenched QCD}
\label{sec:pure_gauge}
Before performing two-flavor QCD simulations at finite temperature, we
carry out a quenched study in order to validate the strategy of
extracting the physics of the $\theta=0$ vacuum from the fixed
topology simulations at finite temperature.
We measure the topological susceptibility at finite temperature using
the method outlined in the previous section and compare the result
with those in the literature obtained with the conventional method of
calculating the variance of the global topological charge 
$\langle Q^2\rangle$.

\subsection{Setup and datasets}
In order to fix the global topological charge throughout the Hybrid
Monte Carlo (HMC) simulation, the JLQCD collaboration introduced 
two extra species of unphysical Wilson fermions with a large negative
mass $-m_0$ \cite{Fukaya:2006vs}.
With the mass of order of the lattice cutoff, the extra degrees of
freedom are irrelevant in the continuum limit.
Since the overlap-Dirac operator $D_{\rm ov}$ is built upon the
hermitian Wilson-Dirac operator $H_W(-m_0)$ as in (\ref{eq:Dov}), a
change of the index of $D_{\rm ov}$ accompanies a zero-crossing of one
of the eigenvalues of $H_W(-m_0)$.
The extra Wilson fermions generate a fermion determinant of the form
$\det[H_W(-m_0)^2]$ and such zero-crossing is prohibited.
We also introduce twisted-mass ghosts to cancel unwanted contributions
from the extra Wilson fermions.
The net effect for the Boltzman weight in the path integral is
\begin{equation}
  \frac{\det [H_W(-m_0)^2]}{\det [H_W (-m_0)^2+\mu^2]},
  \label{eq:DetFukaya}
\end{equation}
where $\mu$ is the twisted mass given to the ghosts.
The eigenmodes of $|H_W(-m_0)|$ above $\mu$ do not contribute to the
Boltzman weight effectively, and only the near-zero modes are affected.
In this work, we chose $\mu=0.2$ in the lattice unit.

The suppression of the near-zero modes of $H_W(-m_0)$ does not spoil
local topological fluctuations that give rise to the topological
susceptibility $\chi_t$.
Indeed, using the same formulation, $\chi_t$ was successfully
calculated and confirmed to be consistent with the expectation from
chiral perturbation theory \cite{Aoki:2007pw}.

\begin{table*}[tb]
  \begin{tabular}{ c  c  c  c  c  c  c }
    \toprule[0.5pt]
    Type & $\beta$ & $a$ (fm) & $T$ (MeV) & $T/T_c$ &  
    $N_{\text{eigenval}}$ & $N_{\text{correlators}}$\\
    \midrule[0.2pt]
    FT & 2.35  & 0.132   & 249 & 0.86 &  106 &  106  \\ 
    FT & 2.40  & 0.123   & 268 & 0.93 &  336 &  336  \\ 
    FT & 2.43  & 0.117   & 281 & 0.97 &  101 &  101  \\ 
    FT & 2.445 & 0.114   & 288 & 1.00 &  424 &  420  \\ 
    FT & 2.45  & 0.113   & 290 & 1.01 &  584 &  584  \\ 
    FT & 2.46  & 0.111   & 295 & 1.02 &  251 &  245  \\ 
    FT & 2.48  & 0.107   & 306 & 1.06 &  420 &  321  \\ 
    FT & 2.50  & 0.104   & 316 & 1.10 &  379 &  218  \\ 
    FT & 2.55  & 0.094   & 348 & 1.20 &  487 &  487  \\ 
    \midrule[0.3pt]
    CT & 2.58 $Q=0$ & 0.099 & 331 & 1.10  & 299  & 235\\
    CT & 2.58 $Q=1$ & 0.099 & 331 & 1.10  & 262  & 257\\
    \bottomrule[0.5pt]
  \end{tabular}
  \caption{
    \label{tab:pureg_param}
    Parameters for the pure gauge simulations on a $24^3\times 6$
    lattice with the topology-fixing extra Wilson fermions 
    (labeled as FT). 
    The global topological charge is fixed to $Q=0$ in these runs.
    Lattice spacing is estimated from the heavy quark potential.
    The number of configurations used for eigenvalue calculations the
    (disconnected) correlator calculations is given in the columns of 
    $N_{\text{eigenval}}$ and $N_{\text{correlators}}$.
    Also listed are the runs without fixing topology (CT) at
    $\beta=2.58$ (with the Iwasaki gauge action).
    The critical temperature for this case is $\sim$300~MeV
    \cite{Iwasaki:1996sn}. 
    The total number configurations for this run is 1,069.
    Out of them, the sectors of global topological charge $|Q|$ = 0
    and 1 are selected for measurements of the disconnected correlators. 
  }
\end{table*}

We generated finite-temperature data in the pure gauge theory on a
lattice of size $24^3\times 6$.
With $N_t=6$, the transition temperature corresponds to the lattice
spacing $a\simeq$ 0.11~fm, which is in the region where the locality of
the overlap-Dirac operator is satisfied with our choice of lattice
actions. 
We take a range of lattice spacing 0.09--0.13~fm, which corresponds to
$\beta$ = 2.35--2.55 with the Iwasaki gauge action.
The lattice spacing is estimated from the heavy quark potential 
measured on independent zero temperature lattices with an input of
$r_0$ = 0.49~fm. 
For each parameter, we accumulated $O$(300--500) configurations
separated by 50--100 HMC trajectories in the trivial topological
sector $Q=0$.
The lattice parameters are summarized in Table~\ref{tab:pureg_param}.

The autocorrelation time of standard thermodynamical observables like
internal energy and the order parameters, the Polyakov loop and chiral 
condensate, are negligible. 
Decorrelation of topology-related quantities, such as the lowest
eigenvalue of the overlap-Dirac operator, is very slow at higher
temperatures, which is a well known effect on fine lattices 
(see for example \cite{DelDebbio:2004xh,Schaefer:2010hu}). 

We estimate the finite temperature phase transition point $T_c$ by
an inflection point of the Polyakov loop, which is more
precise than its susceptibility with our statistics. 
The corresponding $\beta$ value is $\beta=2.445$, which corresponds to
288~MeV.
For each lattice, a value of $T/T_c$ is also listed in
Table~\ref{tab:pureg_param}. 
The precise value for $T_c$ is irrelevant for the conclusions of this
section.

\subsection{Eigenvalue spectral density}
Before analyzing the topological susceptibility, we investigate the
eigenvalue distribution of the overlap-Dirac operator, which is
closely related to the topological susceptibility as we discuss below.
Above the transition temperature the density of near-zero eigenvalues
must vanish as dictated by the Banks-Casher relation
\cite{Banks:1979yr}\footnote{
  Strictly speaking. this is valid only when the chiral condensate is
  an order parameter of the phase transition.
  In the quenched theory, this is not absolutely necessary, but we
  assume it as a working hypothesis in this section.
}.
On the other hand, a sharp peak at very low eigenvalues has been
observed in previous works at or above the critical temperature
\cite{Edwards:1999zm,Bornyakov:2008bg,Schafer:1996hv}.
These were conjectured to be related to a presence of local
topological object, such as an instanton-anti-instanton pair.
If it is the case, these near-zero eigenmodes should have significant
contributions to the topological susceptibility.

Eigenvalues of the overlap-Dirac operator (\ref{eq:Dov}) are
calculated applying the Implicitly Restarted Lanczos algorithm
for a chirally projected operator $P_+ D_{\rm ov}P_+$, where
$P_+=(1+\gamma_5)/2$.
It gives a real part of the eigenvalue of $D_{\rm ov}$, 
$\mathrm{Re} \lambda_{\rm ov}$, and then the complex 
$\lambda_{\rm ov}$ is reconstructed by solving a relation
$|1-\lambda_{\rm ov}/m_0|^2=1$, which is a direct consequence of the
Ginsparg-Wilson relation.
The complex eigenvalue $\lambda_{\rm ov}$ is projected onto the
imaginary axis as
$\lambda\equiv\mathrm{Im}\lambda_{\rm ov}/(1-\mathrm{Re}\lambda_{\rm ov}/2m_0)$.

When applying the overlap operator (\ref{eq:Dov}), we approximate the
sign function using a rational approximation with the Zolotarev
coefficients after subtracting a few lowest-lying eigenmodes of the
kernel operator $H_W(-m_0)$. Taking the degree of the rational 
function to be 16th, the precision of the approximation is kept
better than $10^{-10}$. With this tiny error we confirmed that the 
low-lying eigenvalues of $D_{ov}$ are obtained to 8-digits or better.
The precision of the approximation is kept better than $10^{-7}$.
Further details may be found in \cite{Aoki:2012pma}.

\begin{figure}
  \includegraphics[clip=true, width=0.8\columnwidth]{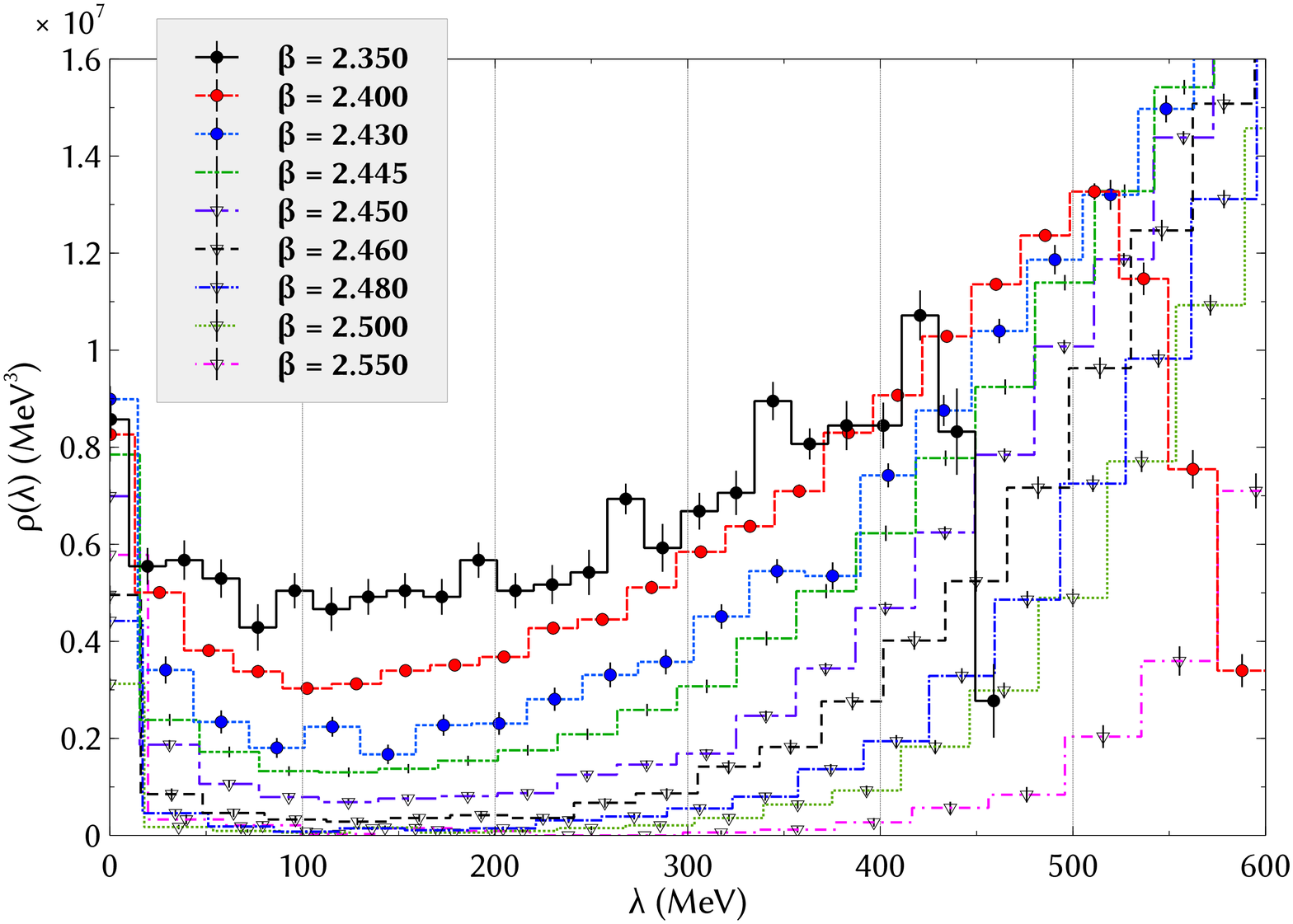}
  \caption{
    \label{fig:Eigenval24}
    Eigenvalue spectral density in quenched QCD at finite temperature.  
    The histograms plotted with filled symbols are those below the
    critical temperature.
    Those above the critical temperature are plotted with open
    symbols.
    The data at {\it would-be} critical point ($\beta=2.445$) are
    shown without symbols.
    The error bars associated with the symbols are estimated using the
    jackknife method.
    The inset shows a magnification of the near-zero mode region,
    showing an accumulation of very low eigenmodes.
  } 
\end{figure}

In Figure~\ref{fig:Eigenval24} we plot the eigenvalue spectral density
observed on pure gauge lattices.
In this study, $\Sigma$ is defined with the lattice regularization and
not converted to the continuum regularization schemes, such as the
$\overline{\mathrm{MS}}$ scheme.
On the quenched lattices, the thermodynamical limit is equivalent to
the infinite volume limit, and our lattices are already sufficiently
close to this limit when extracting $\Sigma$ from the data.
(Note that there is no long-range mode in the pure gauge theory.)

From the plot, we find that at low temperature ($\beta$ = 2.35--2.43)
the eigenvalues accumulate to the amount corresponding to
$\pi\rho(0)\sim$ (200--300~MeV)$^3$.
There is no sharp change observed at the critical temperature
($\beta=2.445$), but $\rho(0)$ decreases towards higher
temperature and eventually vanishes at $\beta=2.55$.

More interestingly, we confirm the presence of the peak near
$\lambda=0$ at around the critical temperature.
Note that our configurations are generated in the $Q=0$ sector and
there is no exact zero modes.
The near-zero modes responsible for the peak have very small
eigenvalues (less than 20~MeV) but they are still non-zero.
We will show that these near-zero modes give the main contribution to
the topological susceptibility.

\subsection{Disconnected correlation functions}
\label{sec:Diconnected_correlation}
So far, very few groups have studied the topological susceptibility
$\chi_t$ at finite temperature in the pure SU(3) gauge theory. 
The most complete results are found in two papers
\cite{Alles:1996nm,Gattringer:2002mr}.
Both calculated the variance of the global topological charge to
obtain $\chi_t=\langle Q^2\rangle/V$.
The former used the geometrical definition of the topological
charge, {\it i.e.} a discretized version of $F_{\mu\nu}\tilde F_{\mu\nu}$,  
on configurations generated with the standard Wilson action.
The topological susceptibility was shown to be stable around the
zero-temperature value ($\sim (180 \text{ MeV})^4$) until the
transition temperature, where it starts decreasing.
A disadvantage of the geometrical method for measuring the topological
charge is that it requires some cooling steps that potentially affect
the final results. 
The other work \cite{Gattringer:2002mr} used the L\"uscher-Weisz
action and directly counted the number of zero modes of a lattice
Dirac operator, which approximately satisfies the Ginsparg-Wilson
relation to measure the global topological charge through the index
theorem. 
The results of these two works reasonably agree with each other.
We take the numbers from \cite{Gattringer:2002mr} for a comparison
with our results.


\begin{figure}
  \includegraphics[clip=true, width=0.8\columnwidth]{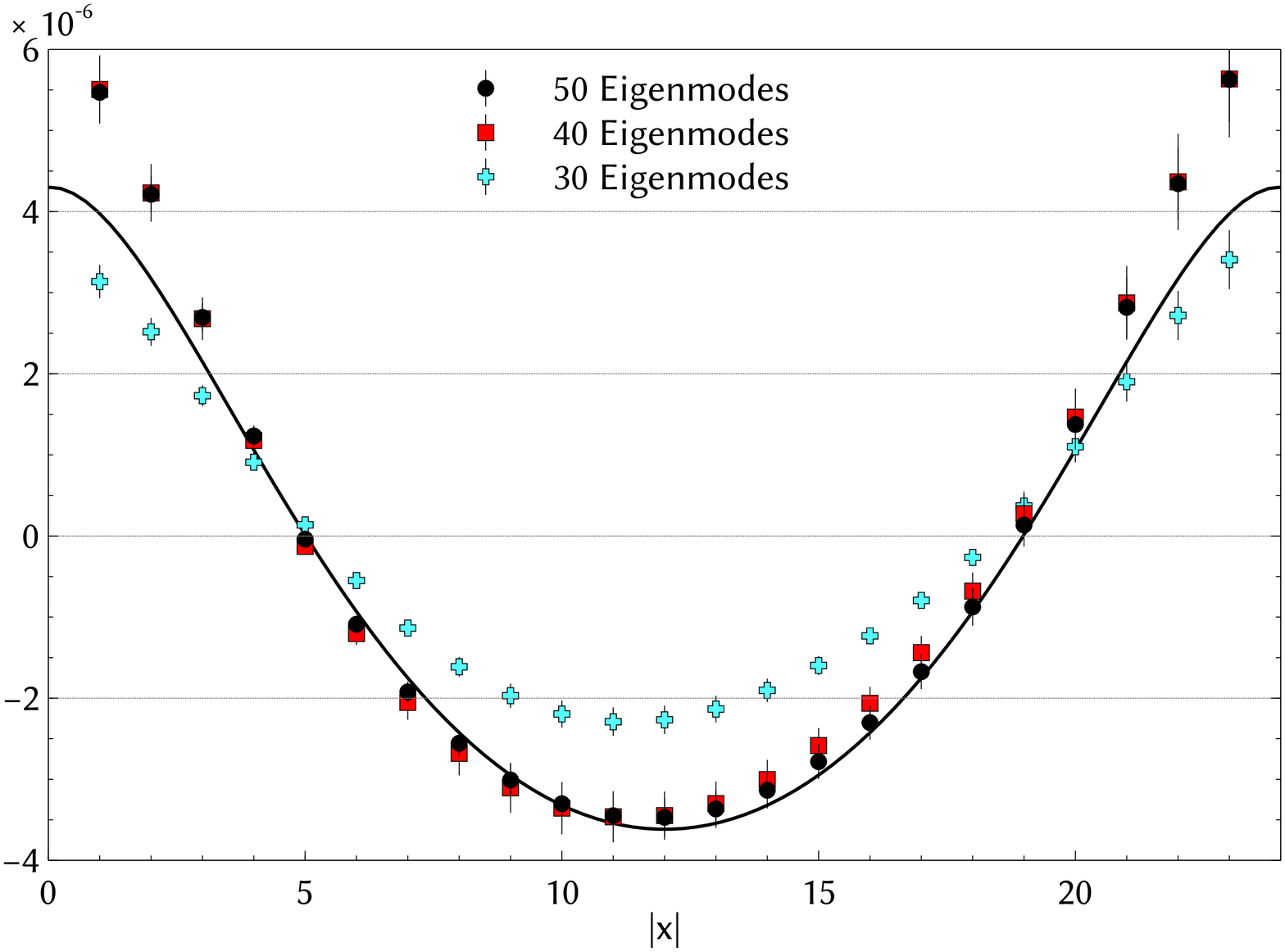}
  \caption{
    Disconnected correlator approximated by different numbers of
    eigenmodes, {\it i.e.} 30, 40 and 50 eigenmodes.
    Data at $\beta=2.40$ and $m=0.05$, which is below the transition
    temperature. The continuous line is a fit of the 50 eigemodes data.
  } 
  \label{fig:NumEigenModes}
\end{figure}

We calculate the topological susceptibility on the gauge
configurations generated at a fixed global topological charge
using the formula (\ref{eq:singlet}). 
It requires a calculation of the disconnected quark-flow diagram,
appearing in the evaluation of the correlation function
$\langle mP(x) mP(0)\rangle$.
Namely, we have to calculate 
\begin{equation}
  \label{eq:disconnected}
  D_{55} (x,0) = 
  \left\langle
    \text{Tr}\left[\gamma_5 S(x,x)\right]
    \text{Tr}\left[\gamma_5 S(0,0)\right]
  \right\rangle
\end{equation}
for each space-time point $x$.
Here, $S(x,y)$ is a quark propagator obtained by an inversion of
the overlap-Dirac operator $D_{\rm ov}$ for a given source point $y$.
Since the numerical cost for a direct calculation is too expensive,
one typically uses some stochastic techniques.
In this work, on the other hand, we introduce a representation of
$S(x,y)$ in terms of the eigenvalues $\lambda_k$ and
eigenvectors $\psi_k(x)$:
\begin{equation}
  S(x,y) = 
  \sum_k \frac{\psi_k(x)\psi_k^\dagger(y)}{\lambda_k + m}.
  \label{eq:SpectralDec}
\end{equation}
This is an exact representation when the sum is taken over all
eigenmodes. 
We truncate the sum to include only the low-lying modes, which should
dominate the long-distance correlations.
We validate this approximation by inspecting the correlation functions
constructed with 30, 40 and 50 low-lying eigenpairs at $\beta=2.40$.
Each eigenpair includes two terms in (\ref{eq:SpectralDec}):
one from the eigenmode of $\lambda_k$ and $\psi_k(x)$, and the other
is that of $\lambda_k^*$ and $\gamma_5\psi_k(x)$, which is also an
eigenmode of $D_{\rm ov}$.
For the valence quark mass $m$ in (\ref{eq:SpectralDec}) we take
$am=0.01$, but the final result for the disconnected diagram is almost
independent of $am$.

Figure~\ref{fig:NumEigenModes} shows the disconnected correlator with 
different numbers of low-lying eigenpairs.
We find that the disconnected correlation function is indeed dominated
by the low-lying eigenmodes and there is no significant difference
between 40 and 50.
Therefore we can safely assume that the evaluation of the disconnected
quark diagram using the 50 eigenpairs is sufficiently precise and
adopt this procedure in the following analysis.


On the quenched lattices, we only evaluate the disconnected quark-line
contributions to the correlator, since the long-distance correlation
from the pion channel does not exist in the disconnected contribution.
Instead, there is a so-called ``hairpin'' contribution 
\cite{Bardeen:2000cz}
of the form
\begin{equation}
  f_P \frac{1}{p^2 + m_\pi^2}m_0^2\frac{1}{p^2 + m_\pi^2}f_P,
\end{equation}
where $m_\pi$ is the pion mass and $f_P$ denotes the matrix element to
annihilate a pion to the vacuum through the pseudo-scalar density
operator. 
The singlet mass-parameter $m_0$ represents a coupling between quark
loops in the quenched vacuum.
In the coordinate space, it corresponds to the functional form
$\sim f_P^2m_0^2(1+m_\pi t)\exp(-m_\pi t)$ 
for zero spatial momentum.
We use this function to fit the lattice data at finite temperature,
taking $t$ in the spatial direction,
together with a constant term $-\chi_t/V$ representing the fixed
topology effect. 
We neglect the subleading effect of $-c_4/(2\chi_tV^2)$ in
(\ref{eq:singlet}).
The pion mass $m_\pi$ is effectively determined by combining a fit of
the connected diagram with that of the disconnected diagram.
The overall coefficient, such as $f_P^2m_0^2$ is treated as a free
parameter. 

An example of the fit is shown in Figure~\ref{fig:NumEigenModes}, which
is for a lattice slightly below the critical temperature.
It demonstrates that the ansatz describes the lattice data well.
The underlying assumption of our analysis is that the pion channel
gives dominant contribution at long distances also at finite
temperature. 
This is a reasonable assumption below the critical temperature $T_c$,
and seems to be valid even above $T_c$ as our data are well fitted.

\subsection{Topological susceptibility}

\begin{figure}[tb]
  \includegraphics[clip=true,width=0.8\columnwidth]{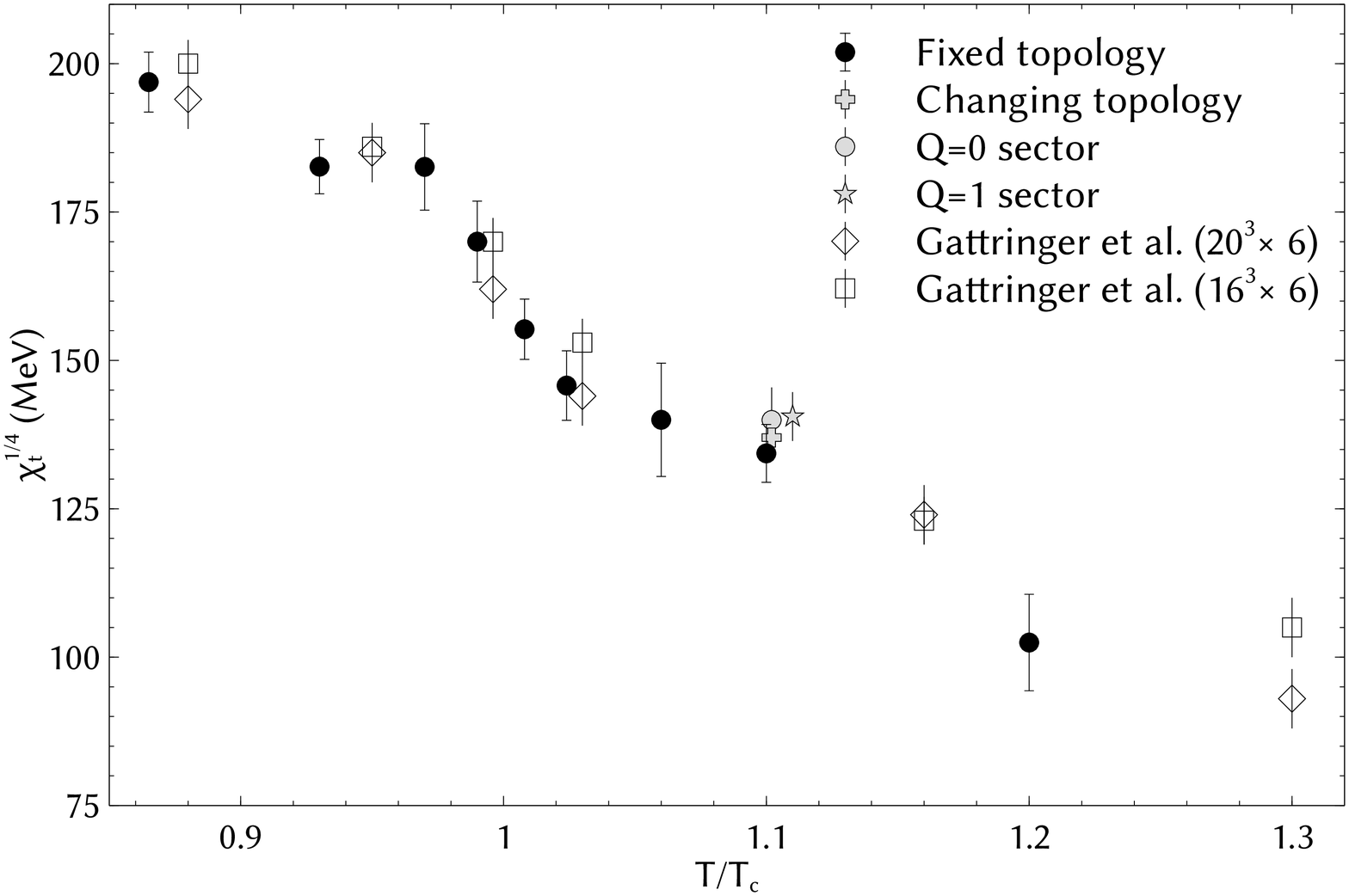}
  \caption{
    Topological susceptibility in quenched QCD calculated on a
    $24^3\times 6$ lattice.
    Data points (black dots) are obtained from a fit of the
    correlation functions (see text).
    The errors are statistical only (jacknife binned). 
    Reference data points from \cite{Gattringer:2002mr} (but on
    $20^3\times 6$ and $16^3\times 6$ lattices) are shown by
    diamonds and squares. The data coming from selected sectors 
    in a changing topology run are also shown ($Q=1$ is slightly
    shifted for readability).
  } 
  \label{fig:TopSusc} 
\end{figure}

The results for the topological susceptibility $\chi_t$ obtained at
a fixed global topology $Q=0$ from the constant long-distance
correlation of the flavor-singlet correlator are shown in
Figure~\ref{fig:TopSusc}. 
Our data are plotted by black dots, which are in good agreement with those from
\cite{Gattringer:2002mr} (squared and diamonds) obtained from the
$Q^2$ distribution.

In order to further cross-check, we also accumulated 1,069
configurations without fixing topology by eliminating the extra Wilson 
fermions (\ref{eq:DetFukaya}).
This run is carried out slightly above the transition temperature by
choosing the lattice spacing $a\simeq$ 0.10~fm.
In Table~\ref{tab:pureg_param}, it is denoted as CT.
The result for $\chi_t$ obtained by counting the number of exact zero
modes is plotted in Figure~\ref{fig:TopSusc} by a cross at around
$T/T_c\simeq 1.1$, which shows a good agreement with our determination
from the fixed topology run.

It is also interesting to see the consistency of this
topology-changing run by selecting configurations of a given $Q$ and
analyzing them with the method for a fixed topology.
We pick up two subsets of configurations with global topological
charge $|Q|$ being 0 and 1, and calculate the disconnected correlation 
function to extract $\chi_t$.
The results are plotted in Figure~\ref{fig:TopSusc} at around
$T/T_c\simeq 1.1$ with different symbols.
They show perfect agreement with the standard method, {\it i.e.}
extracted from $\langle Q^2\rangle/V$, as well as with the result of
the fixed-topology run.
This provides a firm numerical evidence that the method to extract
$\chi_t$ at topology-fixed configurations works as expected, at least for 
$\chi_t > 100$ MeV.

\begin{figure}[tbp]
  \includegraphics[clip=true, width=0.8\columnwidth]{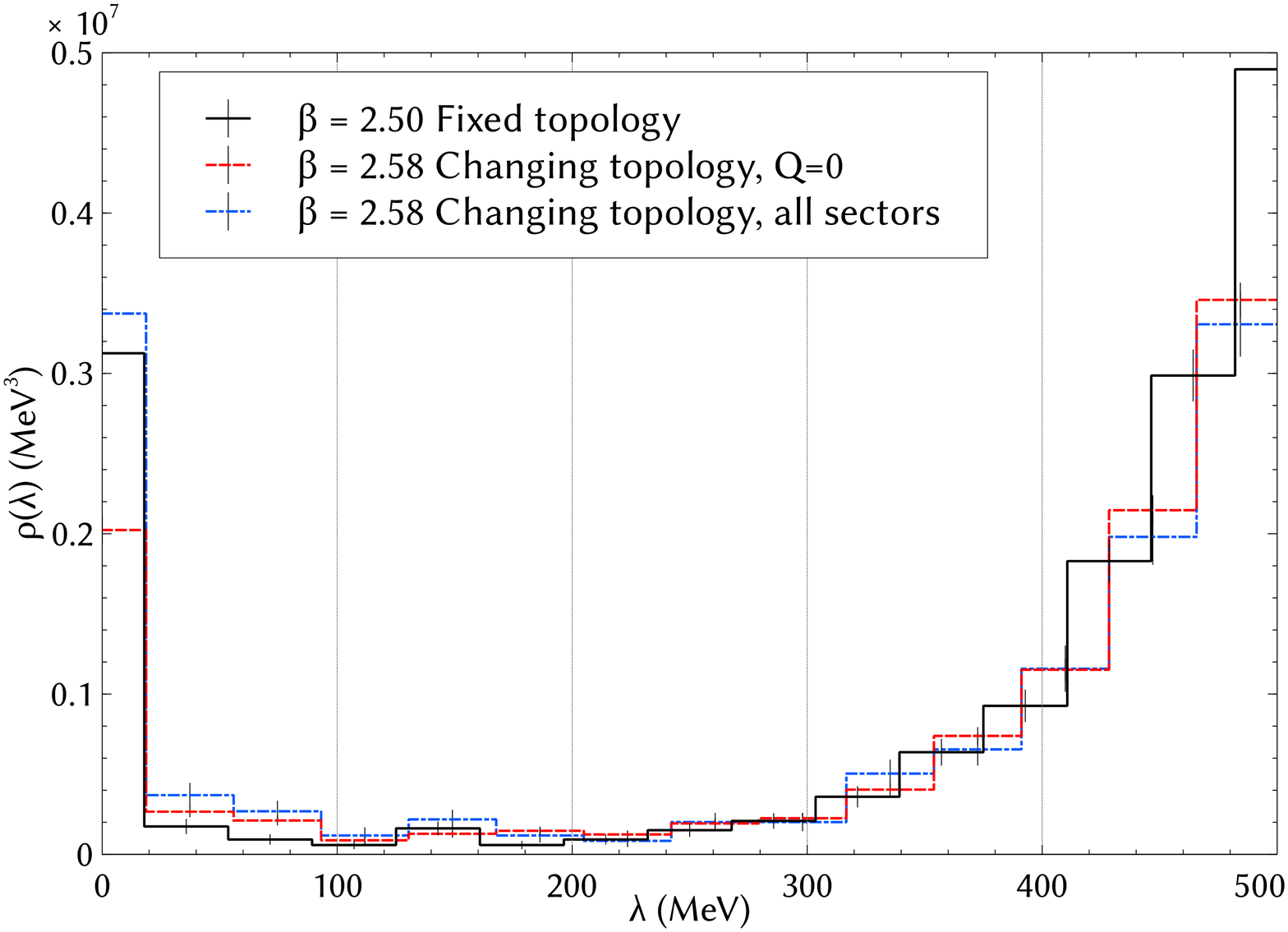}
  \caption{
    Comparison of the eigenvalue spectral density from the runs with and
    without fixing topology at a matched lattice spacing $a\simeq$ 
    0.1~fm.
    The fixed-topology run at $\beta = 2.50$ (black, thin line) is one
    of those shown in Figure~\ref{fig:Eigenval24}. 
    The result of the $Q=0$ configurations selected out of the run
    without fixing topology is overlayed (red, dashed line).
    These two results are in good agreement despite the slight
    mismatch of the lattice spacing (thus temperature).
    The spectral density of all configurations in the 
    topology-changing run (thick line) show a slight deviation at the
    lowest bin. 
    This is understood as an effect of the exact zero modes on the
    $Q\not=0$ configurations, that are taken out in this plot.
    They repel the nearby eigenvalues and make the spectral density lower in
    the vicinity of $\lambda=0$. 
  }
  \label{fig:ComparisonIWrun} 
\end{figure}

We also check that the eigenvalue spectral density does not depend on whether
or not the configurations are generated at a fixed topological sector.
The eigenvalue distribution is compared in
Figure~\ref{fig:ComparisonIWrun}. 
At a matched lattice spacing $a\sim$ 0.1~fm, thus at a matched
temperature $T/T_c\simeq 1.1$, the spectral density at $Q=0$ is
essentially unchanged even when the configuration is generated with
the constraint for the global topological charge.

\section{Dynamical QCD with overlap fermions}
\label{sec:TwoFlav}

Given the theoretical formulation and numerical validations for the
strategy to extract the $\theta=0$ vacuum physics from the
fixed-topology simulations, we embark on a dynamical simulations of
finite temperature QCD using the overlap fermion formulation.
The lattice size in this exploratory study is $16^3\times 8$ and the
global topological charge is fixed to $Q=0$.

\subsection{Run parameters}

\begin{table*}[tbp]
\centering
  \begin{tabular}{ c  c  c  c  c  c  c }
    \toprule[0.5pt]
    $\beta$ & $am$ &  $a$ (fm) & $T$ (MeV) & $T/T_c$ &
    $N_{\text{eigenval}}$ & $N_{\text{correlators}}$\\
    \midrule[0.2pt]
    2.18  & 0.01   & 0.144 & 172 &  0.95  & 118 & 100  \\ 
    2.18  & 0.05   & 0.144 & 172 &  0.95  & 350 & 320  \\ 
    \midrule[0.2pt]
    2.20  & 0.01   & 0.139 & 177 &  0.985 & 187 & 187  \\ 
    2.20  & 0.025  & 0.139 & 177 &  0.985 & 303 & 272  \\ 
    2.20  & 0.05   & 0.139 & 177 &  0.985 & 279 & 279  \\ 
    \midrule[0.2pt]
    2.25  & 0.01   & 0.128 & 192 &  1.06  & 335 & 331  \\ 
    \midrule[0.2pt]
    2.30  & 0.01   & 0.118 & 208 &  1.15  & 512 & 479  \\ 
    2.30  & 0.025  & 0.118 & 208 &  1.15  & 226 & 183  \\ 
    2.30  & 0.05   & 0.118 & 208 &  1.15  & 281 & 281  \\ 
    \midrule[0.2pt]
    2.40  & 0.01   & 0.101 & 243 &  1.35  & 477 & 319  \\ 
    2.40  & 0.05   & 0.101 & 243 &  1.35  & 210 & 210  \\ 
    \midrule[0.2pt]
    2.45  & 0.05   & 0.094 & 262 &  1.45  & 80 & -  \\ 
    \bottomrule[0.5pt]
  \end{tabular}
  \caption{
    \label{tab:DynFerm}
    Parameters for finite temperature QCD simulations with two flavors
    of dynamical overlap fermions.
    The lattice size is $16^3\times 8$.
    The global topology is fixed to $Q=0$.
  }
\end{table*}

In Table \ref{tab:DynFerm} we list simulation parameters. 
We use the Iwasaki gauge action together with the extra Wilson
fermions and associated ghosts (\ref{eq:DetFukaya}).
The temporal size $N_t=8$ is chosen to generate smooth enough gauge 
configurations at around the critical temperature to guarantee the 
locality properties of the overlap operator \cite{Yamada:2006fr}.  
The aspect ratio $L/N_t$ is not sufficiently large for finite
temperature simulation. 
This is a possible source of systematic errors especially in the
vicinity of the phase transition, but we do not consider such errors in
this work, which is the first attempt to extract the physics related
to the axial $U(1)$ sector from the overlap fermion simulations.

Our lattice volume and statistics are not sufficient to precisely
determine the transition point solely from the generated
configurations. 
Instead, we assume a value of the critical temperature 
$T_c$ = 180~MeV in two-flavor QCD. 
The lattice spacing is estimated from an existing analysis at zero
temperature \cite{Aoki:2008tq}.
From its $\beta$ value ($\beta=2.30$), the lattice spacing in the
range ($\beta$ = 2.18--2.45) is obtained assuming the renormalization
group running.
Systematic uncertainty associated with this estimate exists, but it
does not affect the conclusion of this paper, which is qualitative.

The quark mass for degenerate up and down quarks is taken in the range
0.01--0.05 in the lattice unit.
This corresponds to the bare mass of 14--70~MeV for the lattice close
to the transition point ($\beta=2.20$).

\subsection{Dirac operator spectral density}
As in our quenched analysis, we calculate the low-lying eigenvalues of 
the overlap-Dirac operator on two-flavor QCD ensembles.
Since the chiral condensate $\langle\bar{q}q\rangle$ gives an order
parameter of the finite temperature phase transition, according to the
Banks-Casher relation (\ref{eq:Banks-Casher}) the spectral
density near the zero eigenvalue provides a direct measure of the 
phase of the system.
However, we should note that the relation holds only in the
thermodynamical limit ($V\to\infty$ then $m_q\to 0$), and 
the results at finite volume and quark mass have to be taken with care.

For our main interest in this paper, {\it i.e.} the effective
restoration of the axial $U(1)$ symmetry, the spectral density plays
an unique role.
Let us consider the susceptibilities 
$\chi_\delta = \int d^4x \langle j_\delta^a(x) j_\delta^a(0)\rangle$
and
$\chi_\pi = \int d^4x \langle j_\pi^a(x) j_\pi^a(0)\rangle$
of iso-triplet scalar and pseudo-scalar operators
$j_\delta^a(x)=\bar{q}(x)\tau^a q(x)$ and 
$j_\pi^a(x)=\bar{q}(x)\gamma_5\tau^a q(x)$.
($\tau^a$ is the Pauli matrix to specify the isospin component.
We use a somewhat old notation `$\delta$' for the iso-triplet scalar
state. In the modern terminology, it is called $a_0$.)
Using the property that the eigenmodes of the Dirac operator appears
as a complex-conjugate pair of $\lambda_k$ and $\lambda_k^*$ and with
their eigenvectors simply related by $\gamma_5$, 
{\it i.e.} $\psi_k$ and $\gamma_5\psi_k$,
one can show that the difference of the susceptibilities is written
in terms of the eigenvalues:
\begin{equation}
  \label{eq:SuscDiff}
  \chi_\pi - \chi_\delta = 
  \int_0^\infty d\lambda\, \rho(\lambda)
  \frac{4m^2}{(m^2+\lambda^2)^2}.
\end{equation}

Disappearance of $\chi_\pi - \chi_\delta$ suggests
the effective restoration of the axial $U(1)$ above the critical 
temperature, or at least the anomalous violation of the axial $U(1)$ 
cannot be seen in this iso-triplet (pseudo-)scalar channel.

In the broken phase, {\it i.e.} $\rho(0)\ne 0$, the difference \ref{eq:SuscDiff}
diverges as $1/m$ in the chiral limit, 
which is understood as a contribution from the long-distance
correlation due to the Nambu-Goldstone pion channel.
In the symmetric phase $\rho(0)=0$, on the other hand,
the difference survives under the condition  
$\rho(\lambda)\sim\lambda^\alpha$ with $\alpha\leq 1$.
It was recently shown that this condition is not fulfilled
\cite{Aoki:2012yj}, {\it i.e.} $\alpha > 2$.

For other channels, the axial $U(1)$ restoration cannot be simply
parameterized by only using the spectral function and the details of
the eigenvectors are relevant.
We could still expect the important role played by the near-zero
modes, and it is important to identify the strength of its
suppression, {\it i.e.} the power $\alpha$. There is even a
possibility to find a gap in $\rho(\lambda)$, \emph{i.e.} zero density
from $\lambda=0$ up to some value $\lambda_c$.

For this reason, we study the eigenvalue spectral density also in two-flavor
QCD. 
Since it requires the infinite volume limit followed by the
chiral limit, more investigation would be necessary for conclusive results.
The first results obtained in this work with exact chiral symmetry
would still give valuable information towards this goal.

On the configurations generated as listed in Table~\ref{tab:DynFerm},
we calculate 50 lowest eigenvalues and associated eigenvectors of the
hermitian operator $\gamma_5 D_{\rm ov}$.
Paired eigenmodes of $D_{\rm ov}$ can be reconstructed from them, and
we effectively have 100 low-lying eigenmodes.
The numerical method is the same as the one employed in the quenched
study. 
Since the global topological charge is fixed to zero, we do not have
exact zero-modes.

\begin{figure}[tbp]
  \includegraphics[clip=true, width=0.8\columnwidth]{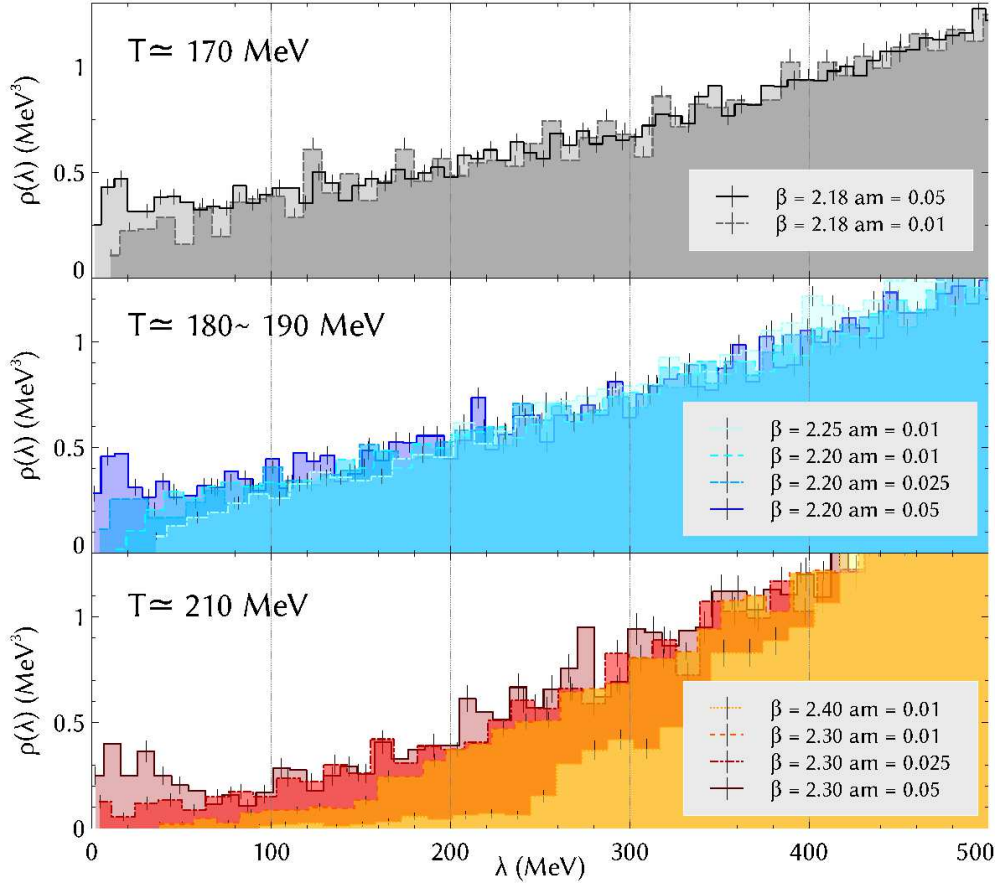}
  \caption{
    Spectral density of the massless overlap-Dirac operator 
    in two-flavor QCD.
    Top and bottom panels are the data clearly below and above the
    critical temperature, respectively.
    The middle panel corresponds to those around the transition
    point. 
    The jackknife errors are shown for each bin of the histogram.
    When the histogram is terminated at the lower end, it implies that
    we find no eigenmode below that value.
    The statistical error in that case is also zero, because we use
    the jackknife method. The lighter the color the lighter the mass.
  } 
  \label{fig:Spect_Dens_Nf2} 
\end{figure}

We plot the results in Figure~\ref{fig:Spect_Dens_Nf2} after
rescaling them to a dimensionful unit to compare the spectra at
different temperature values. 
We investigated the Monte Carlo histories of the lowest mode and found
no evidence to suggest long auto-correlations.
Also, doubling the statistics for a couple of temperatures did not
significantly change the initial result.

First of all, at low temperature ($\beta=2.18$, $T\sim$ 170~MeV, top
panel of Figure~\ref{fig:Spect_Dens_Nf2})
we find significant number of near-zero eigenmodes suggesting
$\Sigma\simeq$ 250~MeV as known in the zero temperature case
\cite{Fukaya:2007fb,Fukaya:2007yv}. 
Such near-zero modes found at a heavier sea quark mass $am=0.05$ are 
suppressed at $am=0.01$.
This is what should happen on a finite volume lattice, even though we
expect non-zero density at $\lambda=0$ in the infinite
volume limit.
Such finite $V$ and finite $m_q$ scalings are studied in detail at
zero temperature
\cite{Fukaya:2007fb,Fukaya:2007yv,Fukaya:2009fh,Fukaya:2010na}.

Near the transition temperature ($\beta=2.20$ and 2.25, $T\sim$
180--190~MeV, middle panel of Figure~\ref{fig:Spect_Dens_Nf2}), 
the result is qualitatively unchanged.
Even above the transition temperature ($\beta=2.30$ and 2.40, $T\sim$
210~MeV, bottom panel of Figure~\ref{fig:Spect_Dens_Nf2}), 
we still see a similar number of near-zero modes when the
quark mass is large ($am=0.05$).
This is consistent with our observation in the quenched theory
(Figure~\ref{fig:Eigenval24}); it indicates that at the quark mass
$am=0.05$ the system is qualitatively similar to the pure gauge
theory. 

Once the quark mass is decreased towards chiral limit at higher
temperatures, not only the near-zero modes, say those below 10~MeV,
but the modes up to $\sim$40~MeV disappears, 
which is very different from what we found on the lower-temperature
lattices ($T\sim$ 170~MeV). 
Our observation suggests the following picture:
In the thermodynamical limit, $\rho(0)$ disappears at the critical
point (by definition).
At a temperature slightly above, the suppression towards the chiral limit
occurs more rapidly and at some higher temperature a gap opens.
This means a stronger suppression than any power $\alpha$ of the form
$\sim\lambda^\alpha$. 
Unfortunately, any quantitative argument about the power $\alpha$ and
the point where gap opens would not be possible with the currently
available data.
There is even a possibility that the gap develops right above the
critical point.
Much more extensive data at several quark masses and volumes would be
necessary for a definite conclusion on this point.

\subsection{Meson correlators}
Further information about the restoration of the symmetry can be 
extracted by directly inspecting the degeneracy of meson correlators.
Under the flavor-singlet (or iso-singlet) axial $U(1)$
transformation, the pion channel is related to $\delta$, likewise the 
$\sigma$ channel is related to $\eta$.
On the other hand, flavor non-singlet (or iso-triplet) chiral
transformation connects $\sigma$ to $\pi$ and $\eta$ to $\delta$.
Therefore, above the transition temperature we expect a degeneracy
between $\sigma$ and $\pi$ as well as between $\eta$ and $\delta$.
If the axial $U(1)$ symmetry is effectively restored, we should see
the degeneracy between $\sigma$ and $\eta$ as well as between $\pi$
and $\delta$.
Namely, all four-channels should become degenerate.

Since the difference between $\pi$ and $\eta$ or between $\sigma$ and
$\delta$ comes from disconnected quark-flow diagram, the $U(1)$
restoration means the absence of the disconnected diagram.
The disconnected contributions to the iso-singlet scalar and
pseudo-scalar correlators are written as
\begin{eqnarray}
  D(x,y) &=& 
  \left\langle
    {\rm Tr}[S(x,x)]\, {\rm Tr}[S(y,y)]
  \right\rangle,
  \\
  D_5(x,y) &=& 
  \left\langle
    {\rm Tr}[\gamma_5S(x,x)]\, {\rm Tr}[\gamma_5S(y,y)]
  \right\rangle,
\end{eqnarray}
where $S(x,y)$ is the quark propagator.
We calculate these disconnected contributions using the eigenmode
decomposition of the quark propagator (\ref{eq:SpectralDec}).
Only the low-lying modes that we calculated for the Dirac operator
spectral density are included in this analysis.
We checked that the correlators at the long distances are
unchanged when the number of low-modes are reduced from 50 to 40. 
Short-distance part is of course largely changed.

\begin{figure}[tbp]
  \includegraphics[clip=true, width=0.45\columnwidth]{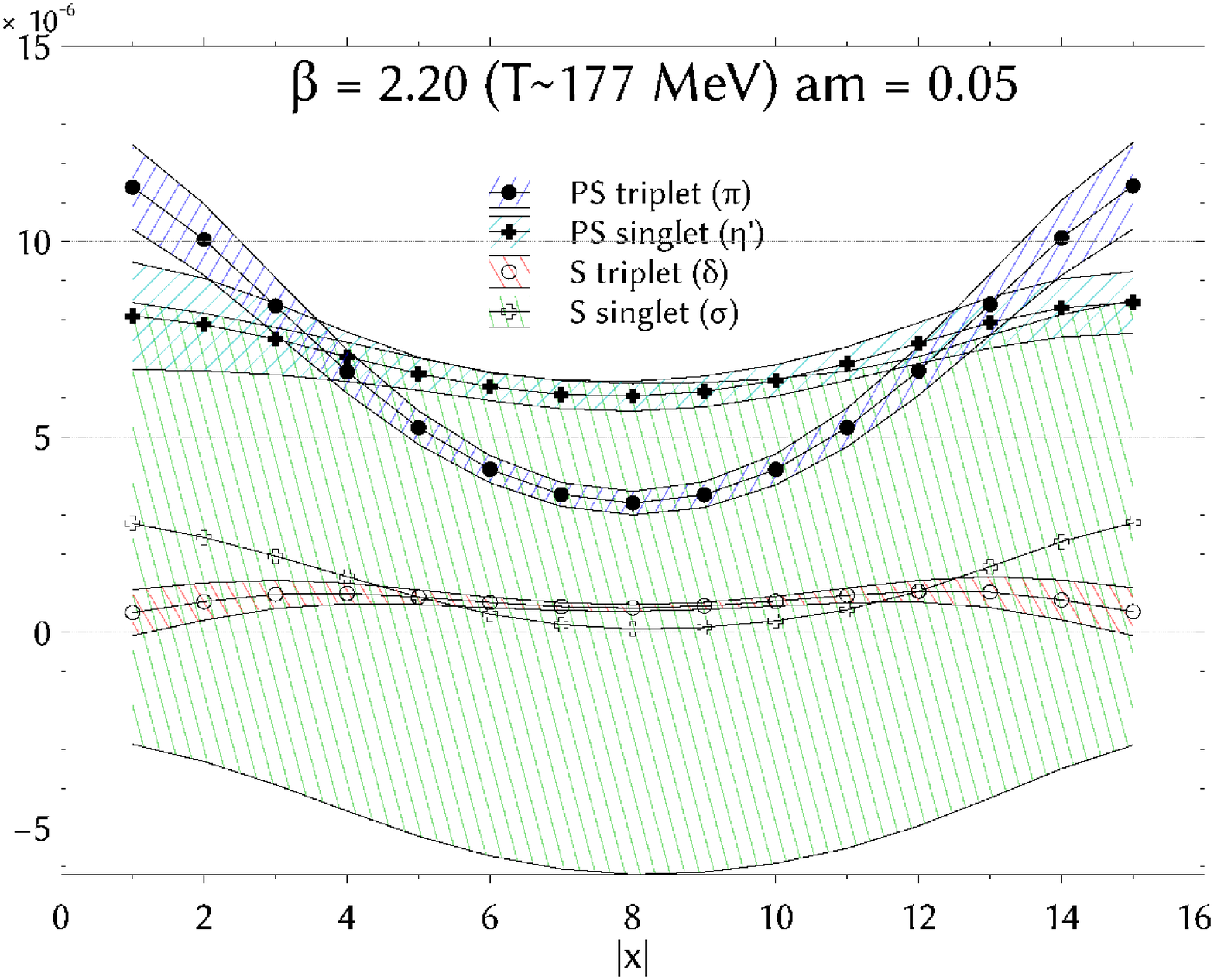}
  \includegraphics[clip=true, width=0.45\columnwidth]{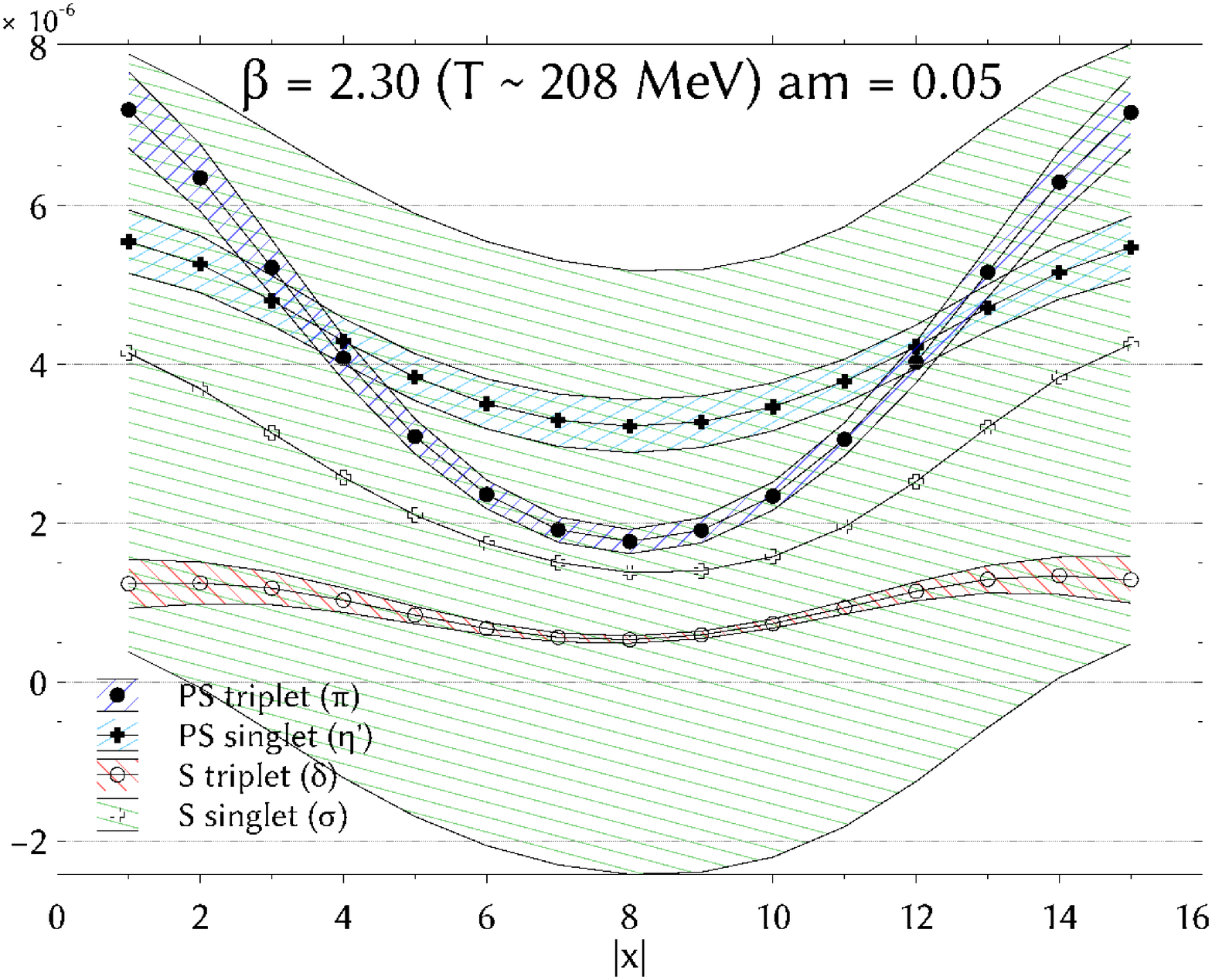}\\
  \includegraphics[clip=true, width=0.45\columnwidth]{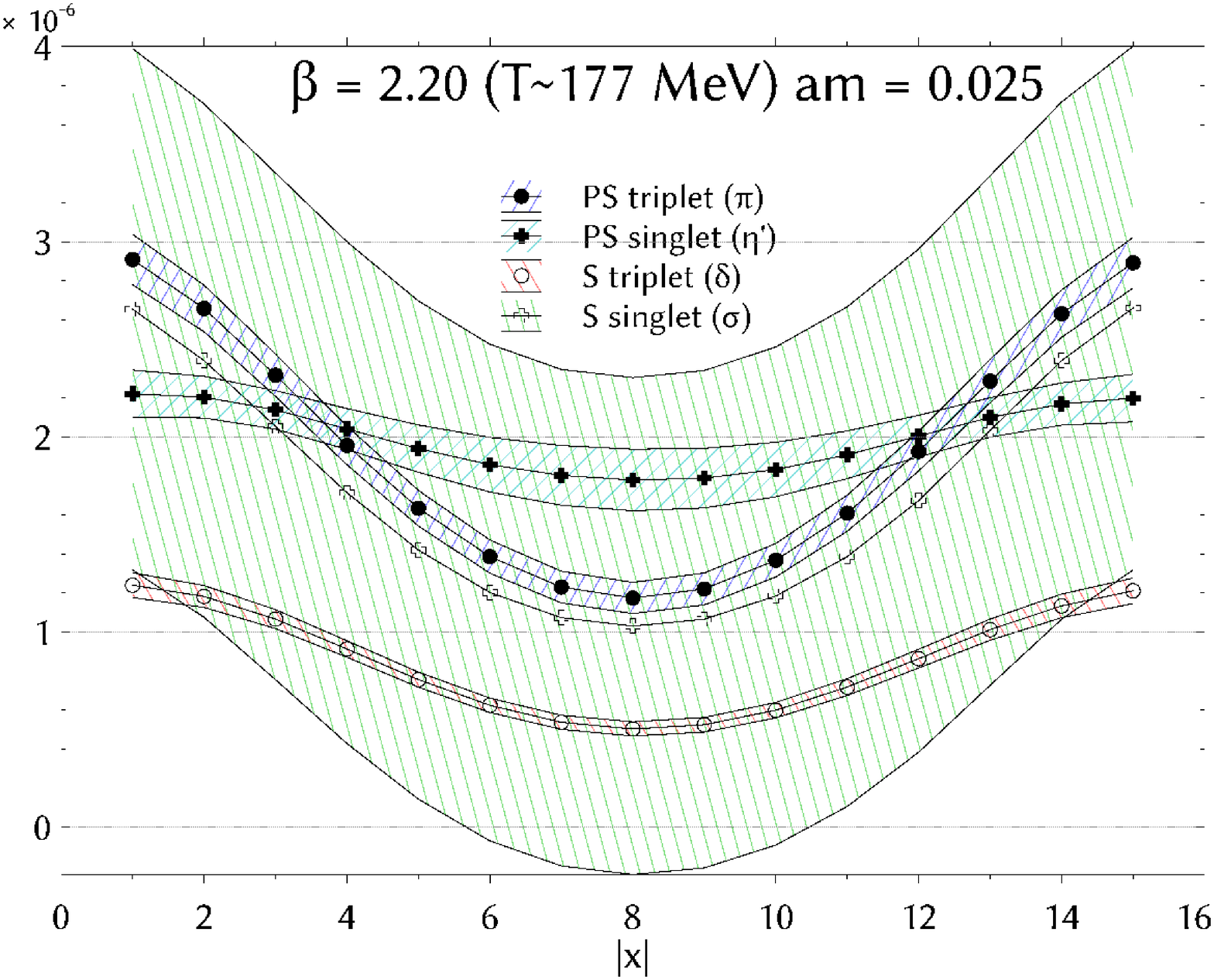}
  \includegraphics[clip=true, width=0.45\columnwidth]{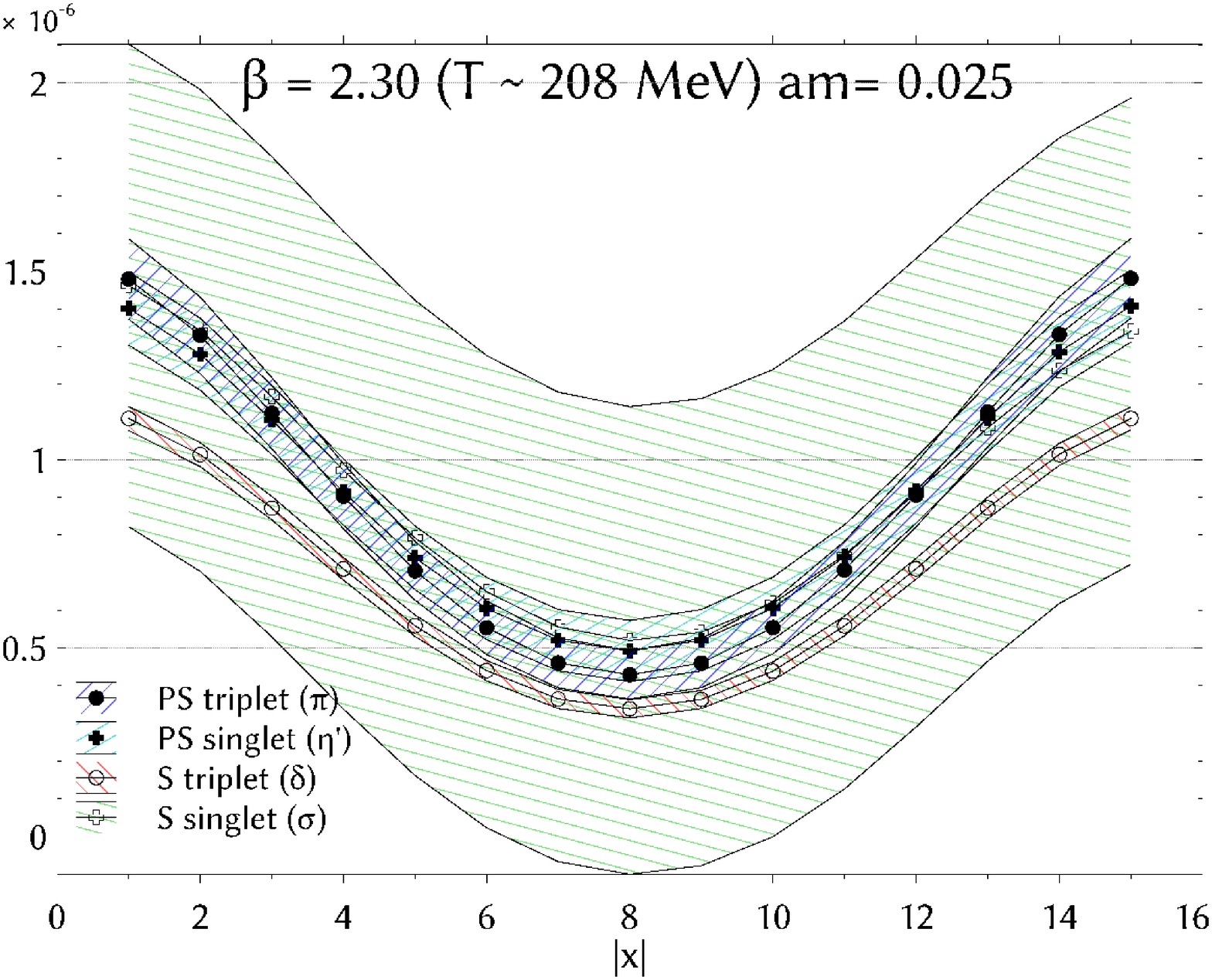}\\
  \includegraphics[clip=true, width=0.45\columnwidth]{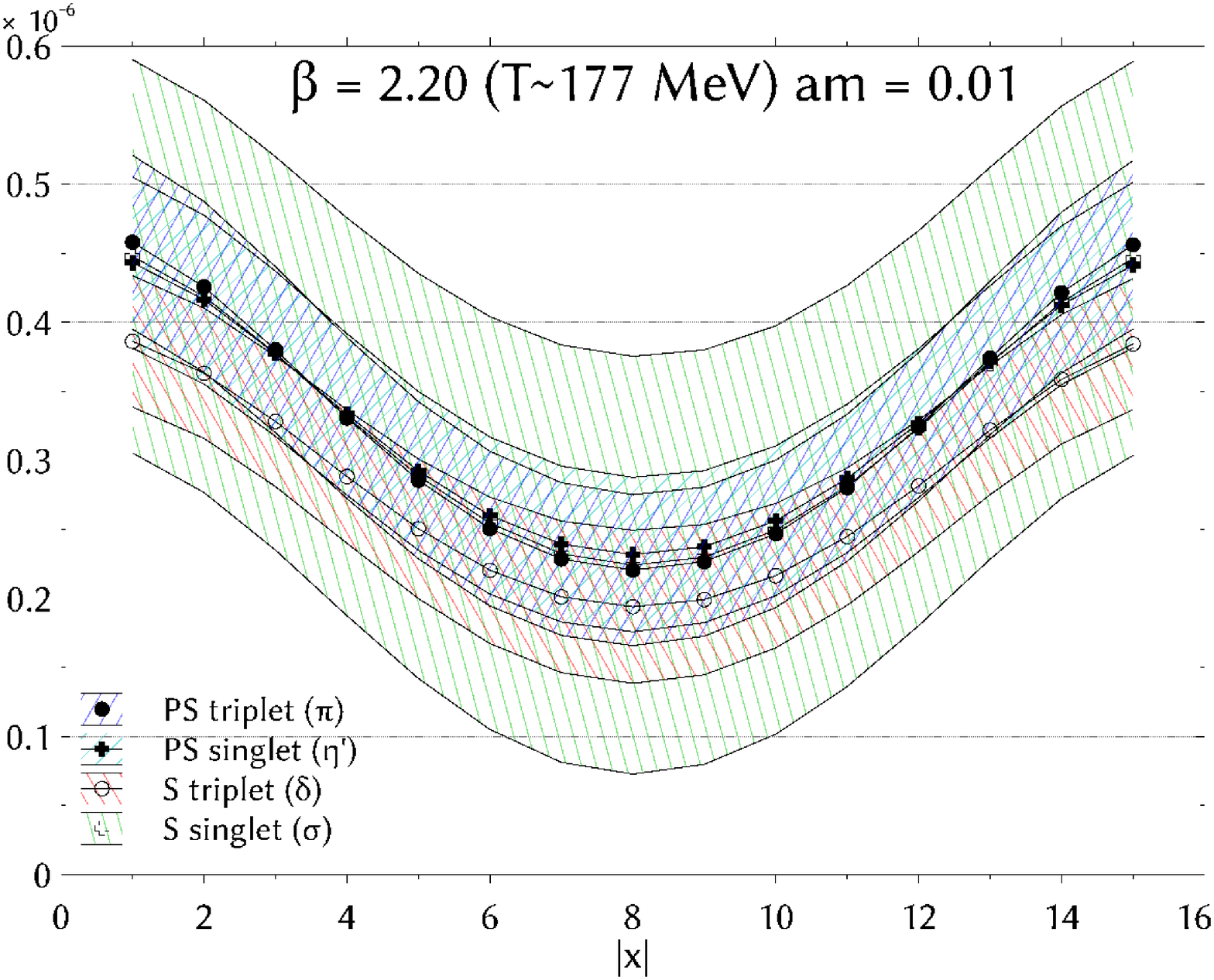}
  \includegraphics[clip=true, width=0.45\columnwidth]{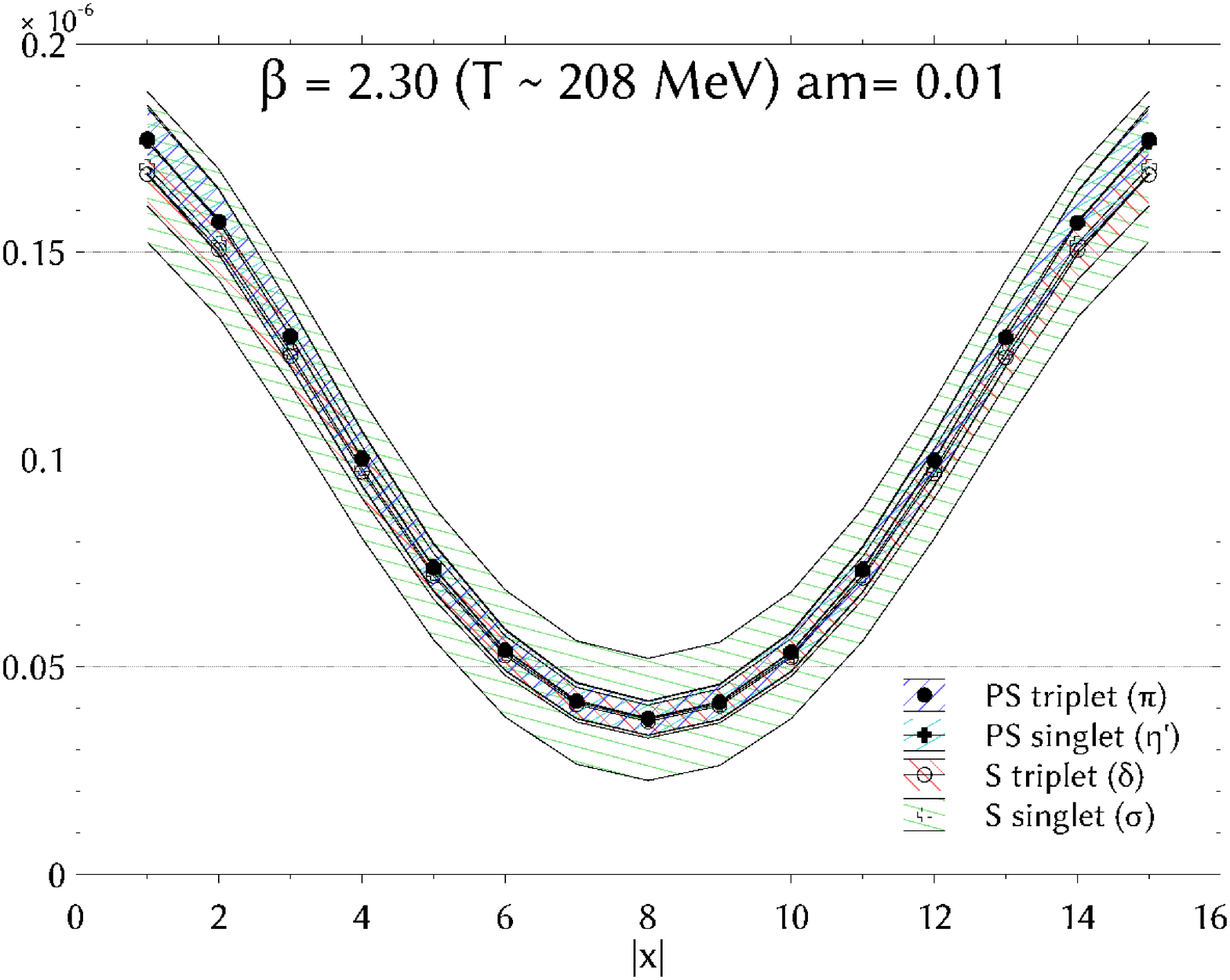}
  \caption{
    Meson correlators at $\beta=2.20$ ($T\simeq$ 180~MeV) and $\beta=2.30$
    ($T\simeq$ 208~MeV). 
    Sea quark masses are $am$ = 0.05, 0.025 and 0.01.
    Results for the $\pi$, $\delta$, $\eta$ and $\sigma$ channels are
    shown.
    Bands represent the statistical error.
  } 
  \label{fig:Mesons_Nf2_220-230} 
\end{figure}



\begin{figure}[tbp]
  \includegraphics[clip=true, width=0.45\columnwidth]{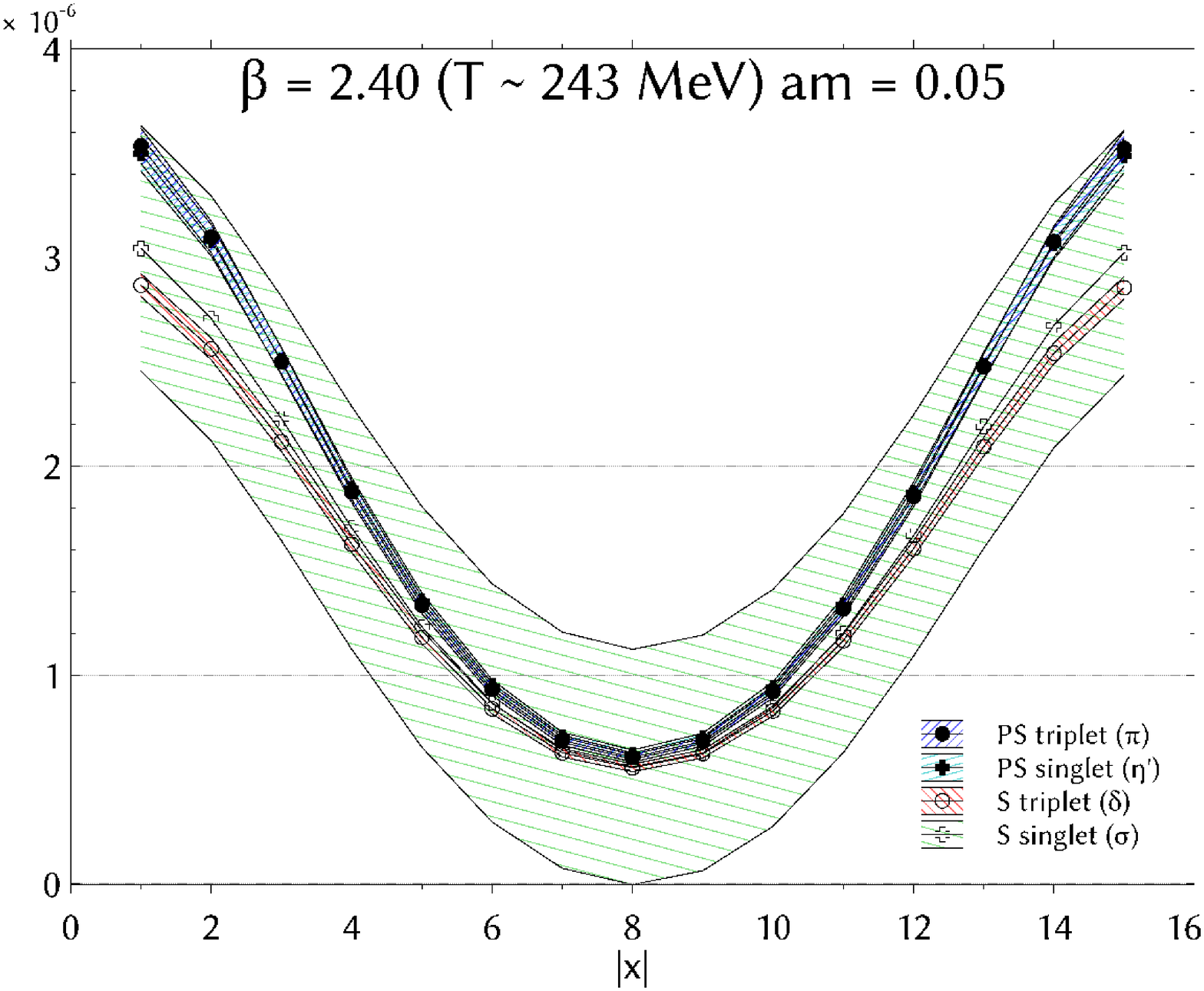}
  \includegraphics[clip=true, width=0.45\columnwidth]{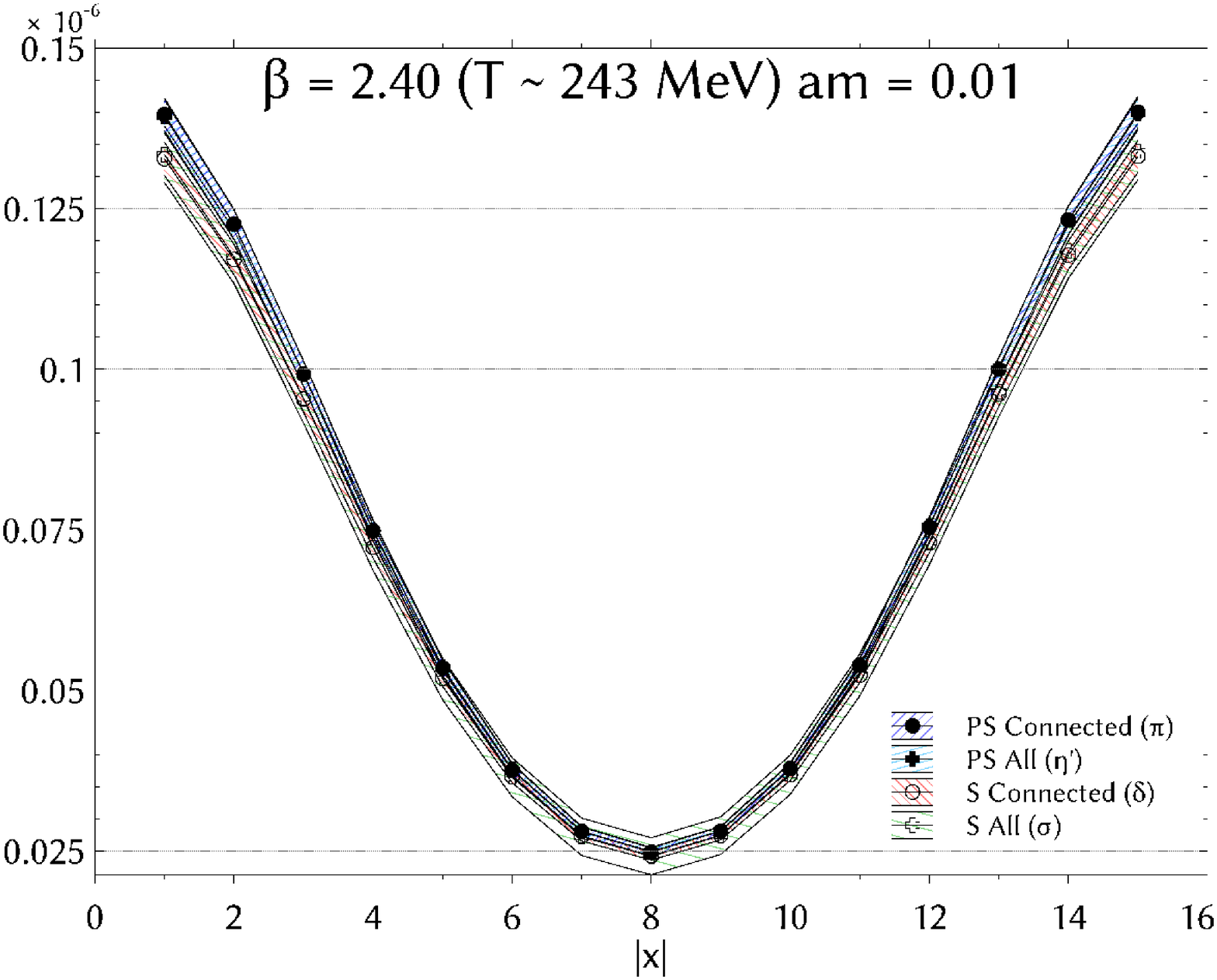}
  \caption{
    Same as Figure~\ref{fig:Mesons_Nf2_220-230}, but for $\beta=2.40$.
  }
  \label{fig:Mesons_Nf2_240} 
\end{figure}

The correlation functions of the relevant mesonic channels for $am=0.05, 0.025$ and 0.01 
are plotted in Figure~\ref{fig:Mesons_Nf2_220-230} from top to bottom panels.
The correlators are for spatial directions and averaged over other
directions. Left panels show the plots at $\beta=2.20$,
slightly below the transition temperature.
We find clear distinctions among different channels at $am$ = 0.05 and
0.025. 
As the chiral limit is approached ($am=0.01$), on the other hand,
all the four channels are nearly degenerate.
Since the disappearance of the near-zero modes becomes significant at
$am=0.01$ (see Figure~\ref{fig:Spect_Dens_Nf2}, middle panel), 
we infer that the difference among the four mesonic channels
indeed originates from the near-zero eigenmodes of the Dirac operator.

Above the transition temperature, $\beta=2.30$ ($T\sim$ 210~MeV), we
find similar degeneracy for small quark masses, as shown in right panels of
Figure~\ref{fig:Mesons_Nf2_220-230}. 
At even higher temperature, $\beta=2.40$ ($T\sim$ 240~MeV), the
degeneracy is found at higher quark masses
(see Figure~\ref{fig:Mesons_Nf2_240}). 
These observations are consistent with our interpretation that the
near-zero eigenmodes below, say, 20~MeV are responsible for the
splitting of the chiral partners for both iso-singlet and iso-triplet
symmetries.

\begin{figure}[tbp]
  \includegraphics[clip=true, width=0.45\columnwidth]{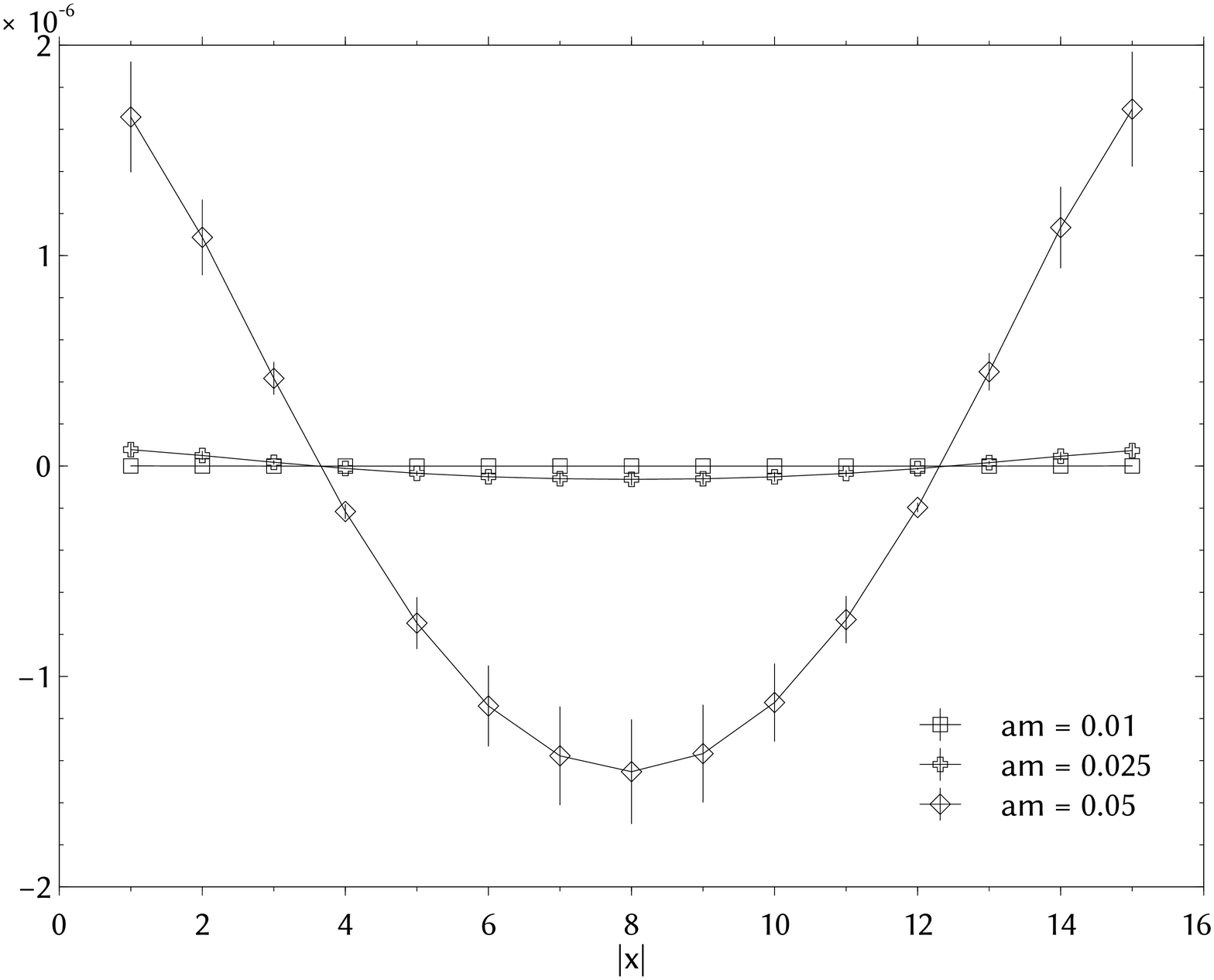}
  \includegraphics[clip=true, width=0.45\columnwidth]{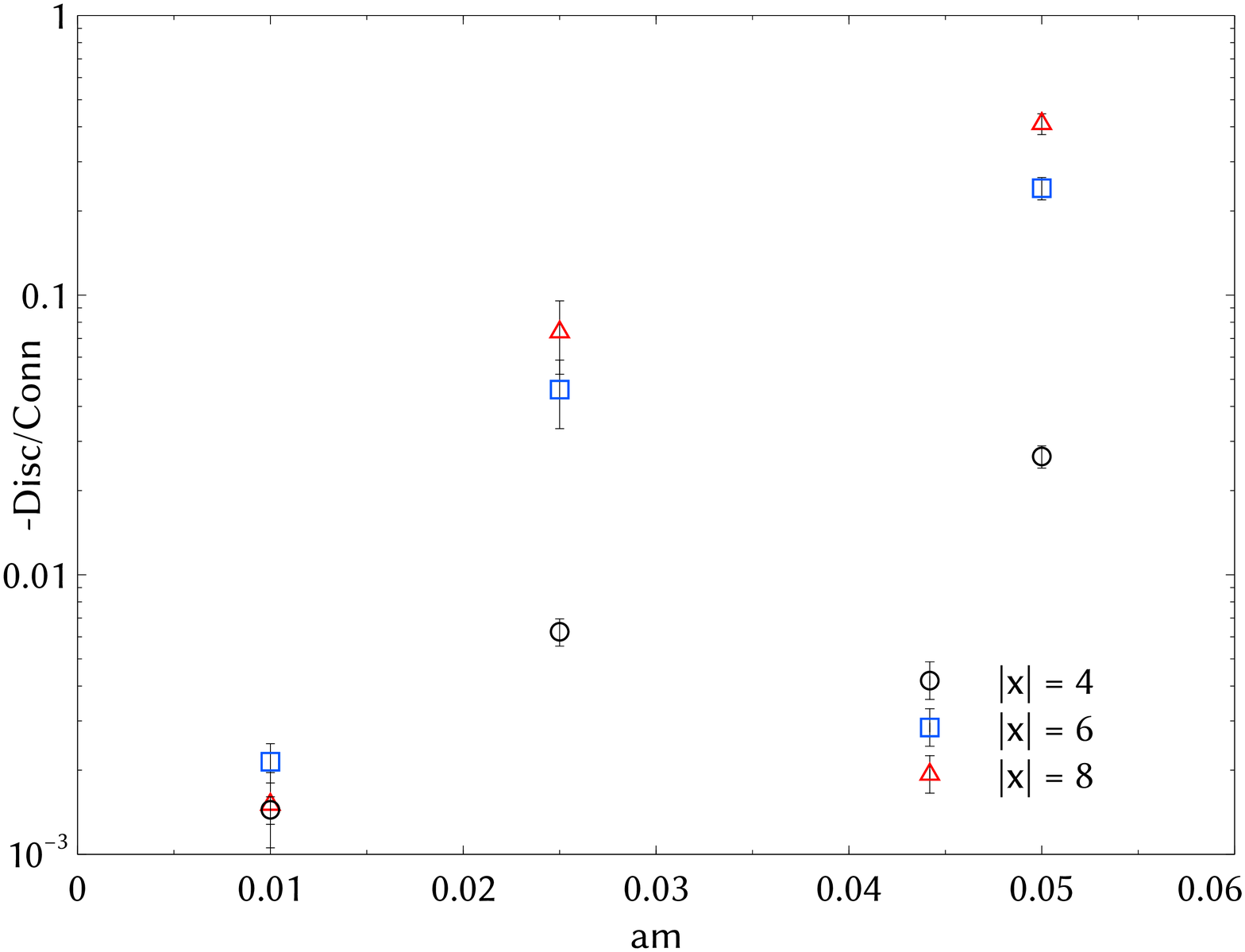}
  \caption{
    (Left panel) Contribution from the disconnected diagram to the iso-singlet
    pseudo-scalar correlator.
    Results at three different quark masses at $\beta=2.30$ are
    plotted. (Right panel) Relative contribution of the disconnected diagram
    to the connected is plotted in a logarithmic scale for some values 
    of $|x|$. It shows that the disconnected contribution rapidly vanishes in 
    the chiral limit.
  }
  \label{fig:Disconnected_PS}
\end{figure}

Figure~\ref{fig:Disconnected_PS} shows the contribution of the
disconnected diagram at $\beta=2.30$.
It is clear from this plot that the disconnected contribution vanishes
at smaller quark masses.

\section{Conclusions}
Finite temperature phase transition of two-flavor QCD could be more
complicated than previously thought due to the possibility of effective
restoration of the axial $U(1)$ symmetry.
The standard analysis \cite{Pisarski:1983ms}
assuming the pattern of symmetry breaking from 
$SU(2)_L\otimes SU(2)_R\otimes U(1)_V$ to $SU(2)_V\otimes U(1)_V$ 
can not be directly applied if the symmetry is effectively extended to
include $U(1)_A$.

This work is one of the first attempts to understand the situation.
Since the axial-anomaly sector of QCD is concerned, the standard
technique of using the staggered fermions in lattice QCD is not
appropriate for this purpose unless one approaches sufficiently close
to the continuum limit.
We instead use the overlap fermion formulation that exactly preserves
chiral symmetry at the classical level and the $U(1)_A$ is violated by
the axial anomaly as in the continuum theory.

A side-effect of using the exactly chiral fermion formulation is that
the global topology has to be fixed in the Monte Carlo simulation.
Theoretical formula have been developed to obtain the correct physics
of the $\theta$ vacuum from such topology-fixed simulations.
Essentially the error due to fixing topology is a finite volume effect
of $O(1/V)$ and the leading $1/V$ correction can be extracted from the
lattice data.
We numerically tested the formula using quenched QCD at finite
temperature, by calculating the topological susceptibility on the
fixed-topology configurations and confirmed that the result agrees
with that of the standard method. 

Having established that fixing topology at finite temperature does not
introduce unexpected systematic effects, we carried out a series of
two-flavor QCD simulations on $16^3\times 8$ lattices around the
transition temperature. 
By measuring the eigenmodes of the overlap-Dirac operator we found
a gap of the spectral density in the chiral limit at high temperature, {\it i.e.} the near-zero
modes disappear not just $\lambda\simeq 0$ but including the modes of
several tens of MeV.
The disappearance of such near-zero modes has a correspondence 
with the degeneracy of meson correlators in the channels related by
the $U(1)_A$ transformation, such as those between $\pi$ and $\delta$
or $\eta$ and $\sigma$.
The origin of this degeneracy is understood:
the near-zero modes are the dominant source of the
disconnected-diagram contributions to the meson correlators at long
distances.

For more quantitative understanding, several sources of systematic errors
are to be addressed in the future works.
Those include the finite volume effect and chiral extrapolation
especially near the transition point.
A cross-checking of the current results with an action that allows
topology change while retaining a very good chiral symmetry would also
be useful.
Along this line, precise study of the order and critical exponents of
finite temperature phase transition of two-flavor QCD, which has not
been established yet, will become feasible taking account of the
effect of axial anomaly.

\section*{Acknowledgments}
Numerical simulations are performed on Hitachi SR11000, SR16000 and
IBM System Blue Gene Solution at KEK under a support of its Large
Scale Simulation Program (No. 09/10-09, 11-05) as well as on Hitachi
SR16000 at YITP in Kyoto University. This work is supported in part by
the Grand-in-Aid of the Japanese Ministry of Education (No. 21674002)
and by the Grant-in-Aid for Scientific Research on Innovative Areas
(No.2004: 23105710, 20105001, 20105003, 20105005) and SPIRE (Strategic
Program for Innovative Research). 

\bibliography{finiteT_ov.bib}

\end{document}